\documentclass[lettersize,journal]{IEEEtran}
\usepackage{amsmath,amsfonts}
\usepackage{algorithmic}
\usepackage{algorithm}
\usepackage{array}
\usepackage[caption=false,font=normalsize,labelfont=sf,textfont=sf]{subfig}
\usepackage{textcomp}
\usepackage{stfloats}
\usepackage{makecell}
\usepackage{threeparttable}
\usepackage{url}
\usepackage{verbatim}
\usepackage{graphicx}
\usepackage{cite}
\usepackage{pifont}
\usepackage{color}
\usepackage{hyperref}
\hypersetup{
	colorlinks=true,
	linkcolor=blue,
	filecolor=blue,      
	urlcolor=blue,
	citecolor=blue,
}
\begin{document}

\title{Advancing LLM-Based Security Automation with Customized Group Relative Policy Optimization for Zero-Touch Networks}

\author{Xinye Cao,\IEEEmembership{~Graduate Student Member,~IEEE,} Yihan Lin, Guoshun Nan,\IEEEmembership{~Member,~IEEE,} Qinchuan Zhou,\\ Yuhang Luo, Yurui Gao, Zeliang Zhang, Haolang Lu, Qimei Cui,\IEEEmembership{~Senior Member,~IEEE,} \\Yanzhao Hou,\IEEEmembership{~Member,~IEEE,} Xiaofeng Tao,\IEEEmembership{~Senior Member,~IEEE,} Tony Q.S. Quek,\IEEEmembership{~Fellow,~IEEE}
\thanks{This work was supported in part by the National Natural Science Foundation of China under Grant 62471064; in part by the National Research Foundation, Singapore and Infocomm Media Development Authority under its Communications and Connectivity Bridging Funding Initiative; in part by the Beijing Natural Science Foundation Program (No.L232002); in part by Beijing University of Posts and Telecommunications (BUPT) Excellent Ph. D. Students Foundation under Grant CX20252013; and in part by Beijing Natural Science Foundation under Grant QY25332. Any opinions, findings and conclusions or recommendations expressed in this material are those of the author(s) and do not reflect the views of National Research Foundation, Singapore. \textit{(Xinye Cao and Yihan Lin contributed equally to this work.) (Corresponding authors: Guoshun Nan; Tony Q.S. Quek.)}}
\thanks{Xinye Cao, Yihan Lin, Guoshun Nan, Qinchuan Zhou, Yuhang Luo, Yurui Gao, Zeliang Zhang, Haolang Lu, Qimei Cui, Yanzhao Hou, Xiaofeng Tao are with National Engineering Research Center for Mobile Network Technologies, Beijing University of Posts and Telecommunications, China. (e-mail: caoxinye@bupt.edu.cn; linjhs@bupt.edu.cn; nanguo2021@bupt.edu.cn; kevinlvrain@bupt.edu.cn; lyh\_eddiemurphy@bupt.edu.cn; gaoyurui813@bupt.edu.cn; 2023211490@bupt.cn; lhl\_2507@bupt.edu.cn; cuiqimei@bupt.edu.cn; houyanzhao@bupt.edu.cn; taoxf.bupt@gmail.com).}
\thanks{Tony Q.S. Quek is with the Singapore University of Technology and Design, Singapore 487372, and also with the Department of Electronic Engineering, Kyung Hee University, Yongin 17104, South Korea (e-mail: tonyquek@sutd.edu.sg).}
}

\markboth{Journal of \LaTeX\ Class Files,~Vol.~14, No.~8, August~2021}
{Shell \MakeLowercase{\textit{et al.}}: A Sample Article Using IEEEtran.cls for IEEE Journals}

\maketitle

\begin{abstract}
Zero-Touch Networks (ZTNs) represent a transformative paradigm toward fully automated and intelligent network management, providing the scalability and adaptability required for the complexity of sixth-generation (6G) networks. However, the distributed architecture, high openness, and deep heterogeneity of 6G networks expand the attack surface and pose unprecedented security challenges. To address this, security automation aims to enable intelligent security management across dynamic and complex environments, serving as a key capability for securing 6G ZTNs. Despite its promise, implementing security automation in 6G ZTNs presents two primary challenges: 1) automating the lifecycle from security strategy generation to validation and update under real-world, parallel, and adversarial conditions, and 2) adapting security strategies to evolving threats and dynamic environments. This motivates us to propose SecLoop and SA-GRPO. SecLoop constitutes the first fully automated framework that integrates large language models (LLMs) across the entire lifecycle of security strategy generation, orchestration, response, and feedback, enabling intelligent and adaptive defenses in dynamic network environments, thus tackling the first challenge. Furthermore, we propose SA-GRPO, a novel security-aware group relative policy optimization algorithm that iteratively refines security strategies by contrasting group feedback collected from parallel SecLoop executions, thereby addressing the second challenge. Extensive real-world experiments on five benchmarks, including 11 MITRE ATT\&CK processes and over 20 types of attacks, demonstrate the superiority of the proposed SecLoop and SA-GRPO. We will release our platform to the community, facilitating the advancement of security automation towards next generation communications.
\end{abstract}


\section{Introduction}
\subsection{Background}

\IEEEPARstart{W}{ith} the commercialization of 5G, research efforts have rapidly shifted toward the exploration of 6G networks. According to the 6G vision recommendation~\cite{wang2023road} released by the ITU-R, security has been identified as one of the fundamental design principles of 6G networks, while the deep integration of artificial intelligence (AI) and communication~\cite{jiang2025comprehensive,zhu2025wireless,cui2025overview} is considered among the six representative usage scenarios. Zero-touch networks~\cite{yang2025towards,masaracchia2023digital,el2024zero} have emerged as a key solution for achieving fully automated network operations, offering essential capabilities such as self-configuration, self-monitoring, self-healing, and self-optimization. ZTNs are expected to play a pivotal role in 6G networks by addressing the growing demand for virtualized network functions~\cite{wang2017computation} and aligning with the trend toward software-defined and automated architectures~\cite{wang2016joint}. The primary objective of ZTNs is to execute various network management and control tasks autonomously, without the need for human intervention.

\begin{figure}[!t]
\centering
\includegraphics[width=0.5\textwidth]{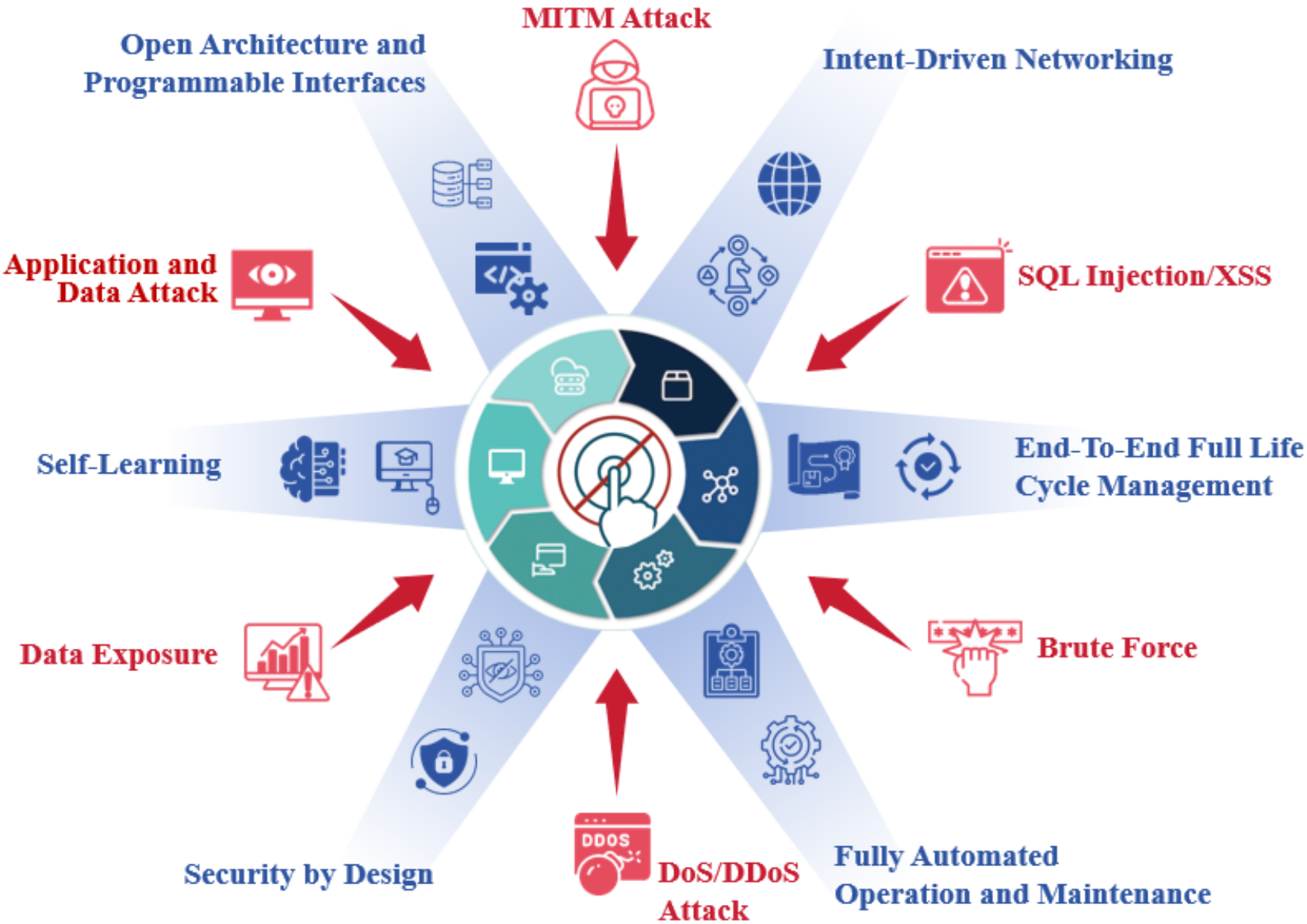}
\caption{Illustration of various attacks in zero-touch networks. Zero-touch networks introduce the software defined network (SDN) framework and automated management. The openness of 6G ZTN is accompanied with various attacks, such as DDoS, SQL injection, and man-in-the-middle (MITM) attacks.}
\label{fig:1_intro}
\end{figure}

6G ZTNs, as illustrated in Figure \ref{fig:1_intro}, expand the attack surface due to the open architectures, distributed networks, and heterogeneous environments, thereby increasing the complexity of threat detection and mitigation~\cite{nguyen2021security,chen2024survey}. For instance, adversaries may launch distributed denial-of-service (DDoS) attacks to disrupt communication services, or SQL injections to affect sensitive user information~\cite{salim2020distributed}. Traditional protection mechanisms~\cite{abasi2023survey,ferrag2023edge} are insufficient to cope with these emerging threats, highlighting the urgent need for intelligent, adaptive, and context-aware security solutions to ensure resilient and efficient defense in next generation communication systems. LLMs~\cite{cao2025exploring,ferrag2025securefalcon,jiang2024large,ferrag2023revolutionizing,jiang2024personalized}, as a transformative technology in the field of artificial intelligence, are reshaping the landscape of network security. With their advanced natural language understanding and autonomous decision-making capabilities, LLMs significantly enhance the effectiveness of threat detection, analysis, and mitigation in next generation communication systems.

\subsection{Motivation} 
The security vulnerabilities exposed by the openness and heterogeneity of 6G and ZTNs call for a fundamental rethinking of network security architecture to defend diverse and evolving threats in dynamic environments~\cite{zheng2022dynamic,khan2024adversarial}. Current security orchestration, automation, and response (SOAR) platforms~\cite{wu2024ontocsd,bartwal2022security,chahal2022proactive} largely rely on handcrafted rule sets or template-based response strategies. However, such approaches often fail to cope with advanced persistent threats, multi-stage attacks, or highly obfuscated adversarial behaviors. Furthermore, most systems only generate recommended strategies, lacking direct executability and seamless integration with diverse security tools, hindering practical deployment. In summary, existing security approaches for 6G ZTNs face two fundamental challenges:

\noindent
\textbf{1) Automating the lifecycle from security strategy generation to validation and update:} Conventional security automation systems are often limited to strategy generation~\cite{liu2024generic,enoch2022practical,wu2024ontocsd} or tool invocation~\cite{sworna2023apiro,sworna2023irp2api}, lacking a complete and automated workflow. They require substantial manual intervention and offer limited adaptability. In contrast, a fully autonomous end-to-end security system must seamlessly coordinate attack simulation, environment configuration, strategy generation, tool execution, and feedback-driven refinement. Achieving such integration poses significant challenges for the intelligence of the system in the real world.

\noindent
\textbf{2) Adapting security strategies to evolving threats and dynamic environments:} Traditional approaches often rely on supervised learning~\cite{yan2025secure,abdallah2022intrusion,faker2019intrusion}, which demands experts to annotate data that is costly and difficult to maintain. Moreover, static datasets fail to capture the dynamic characteristics of real-world attacks, leading to poor generalization for evolving threats and zero-day attacks. To enable adaptive and robust defense, it is essential to develop learning mechanisms with minimal supervision and continuously refine strategies based on real-time feedback from heterogeneous environments.

\subsection{Our Method}
The aforementioned issues motivate us to propose SecLoop and SA-GRPO. A group of security strategies generated by SA-GRPO is deployed across parallel real-world environments instantiated by SecLoop and is iteratively refined based on feedback. We outline six high-level design principles for SecLoop and SA-GRPO to tackle the two challenges.

\noindent
\textbf{1) Learnable:} The system should possess the ability to continuously learn from diverse and evolving network conditions. 

\noindent
\textbf{2) Adaptive:} To ensure resilience in rapidly changing environments, the system must dynamically respond to diverse threat intelligence and adjust defensive strategies accordingly.

\noindent
\textbf{3) Practical:} Generated strategies should be practical and tightly aligned with the execution capabilities of the underlying defense infrastructure.

\noindent
\textbf{4) Automatic:} The system must support full automation across the security lifecycle, from threat simulation to strategy execution and feedback refinement, eliminating the need for manual intervention.

\noindent
\textbf{5) Efficient:} To accelerate learning and improve responsiveness, the system should support parallel execution of candidate strategies in realistic environments, enabling fast validation and feedback cycles.

\noindent
\textbf{6) Pluggable:} Given the heterogeneity of network infrastructures and evolving security tools, the system should maintain a modular architecture, allowing flexible integration and replacement of components.

Keeping the above goals in mind, we design and implement SecLoop, an end-to-end security automation framework that enables strategy generation, execution, and feedback across real-world environments. SecLoop supports parallel deployment of diverse security strategies in isolated, virtualized environments and integrates LLMs as intelligent decision agents. These agents interact with streaming alerts from intrusion detection systems (IDS) and trigger responses through orchestrated security tools. To optimize decision-making under limited supervision, we further propose SA-GRPO, a security-aware group relative policy optimization algorithm. SA-GRPO generates a group of candidate strategies and deploys them concurrently in SecLoop environments. Feedback collected during execution is used to iteratively refine the policy through reinforcement learning. Through customized rewards, including format, execution, evaluation, and penalty, SA-GRPO can be tailored to security-specific scenarios. Extensive experiments on five benchmarks over 20 types of attack demonstrate the effectiveness of our proposed framework. Our code is publicly available\footnote{https://github.com/caoxinye/SecLoop}.

\subsection{Main Contributions}
The main contributions of this paper are threefold:

\noindent
\ding{182} \textbf{LLM Agent Empowered Security System:} We design and implement SecLoop, the first end-to-end security strategy generation, orchestration, response, and feedback system, enabling adaptation of defensive strategies against evolving cyberthreats in 6G zero-touch networks. Integrated with native LLM and automated real-world BATTLE-FIELD (Blue And red Team Tactical Learning Environments For Intrusion, ExpLoitation, and Defense), SecLoop supports comprehensive attack simulation encompassing the ATT\&CK process, provides a robust environment for fine-tuning LLMs and validating security algorithms under realistic adversarial conditions.

\noindent
\ding{183} \textbf{Security-Aware GRPO:} We propose SA-GRPO, a security-aware group relative policy optimization algorithm that refines security strategies through iterative feedback from parallel BATTLE-FIELD. SA-GRPO eliminates the need for high-quality labeled data, adapting to the real world. To further tailor the optimization process to security-specific tasks, we design a customized reward function from four complementary perspectives, including format check, simulation execution, attack evaluation, and reasoning verification.

\noindent
\ding{184} \textbf{Extensive Experiments:} We conduct extensive experiments on four public benchmarks to show the effectiveness of the proposed SA-GRPO, yielding a state-of-the-art defense for the next generation networks. We also build a more comprehensive dataset on the SecLoop, and such a dataset can serve as a benchmark for security orchestration. Furthermore, we conduct 21 types of cyberattacks, including advanced and zero-day attacks, on real-world tests to demonstrate the practical potential of SA-GRPO. Finally, we provide three case studies of heterogeneous edge devices to visually demonstrate the detailed work procedure of the proposed SA-GRPO.

\subsection{Related Work}

\subsubsection{Security Orchestration}
Security orchestration has been widely applied in programmable network architectures such as software-defined networking (SDN)~\cite{dungarani2024sdn,zarca2018managing} and network functions virtualization (NFV)~\cite{robles2025dynamic,basile2019adding} to achieve automated and scalable security management. The zero-touch network and service management (ZSM) proposed by ETSI is regarded as a new paradigm to achieve fully automated network management~\cite{yang2025towards}. In this framework, the security orchestration center (SOC) is deeply integrated to support real-time policy enforcement, intent analysis, and dynamic deployment of cross-domain security services~\cite{batewela2025security,el2024zero,yang2025towards}. The existing research also systematically explores the convergence trends and security challenges of key technologies such as SDN, NFV, Multi-Access Edge Computing (MEC), and O-RAN in 5G and B5G~\cite{batewela2025security,parvez2018survey}. Although SOC provides a complete automated management framework, current research still mostly focuses on the automation of the deployment of security functions~\cite{bohme2024software}. To the best of our knowledge, we are the first to build a fully automated end-to-end security automation framework with native LLMs.

\subsubsection{Security Strategy Optimization}
Reinforcement learning (RL) has been widely adopted for optimizing security strategies. Q-learning~\cite{yu2023reinforcement} enables adaptive threshold tuning for replay attack mitigation, and Double Q-learning~\cite{issa2024cyberrl} enhances stability in edge-based intrusion detection via CyberRL. DQN~\cite{zhang2023game} was applied in the APT Rivalry Game framework to optimize the timing of defense responses. PPO~\cite{goel2024optimizing} further supports attack path planning in dynamic Active Directory graphs, while defenders optimize edge-blocking decisions using value-based evaluations. Hierarchical approaches such as HMARL employ Q-Tabular and PPO~\cite{tang2024method} across different layers to enable multi-agent collaborative defense. 
SAC~\cite{jiang2024rudolf} has been used in the RUDOLF framework to learn adaptive traffic obfuscation strategies in Tor, and DDPG-MIX~\cite{tong2020finding} has been adopted in Double Oracle games to compute robust alert prioritization under adversarial conditions. Despite recent progress, existing RL-based approaches often rely on static datasets and struggle to generalize or adapt in complex environments. To address this, we propose SA-GRPO, a security-aware RL algorithm that iteratively refines policies by contrasting feedback from parallel executions of real-world testbeds, enabling more adaptive strategy generation.

\subsection{Paper Organization and Notations}
The remainder of the paper is organized as follows. Section \ref{2system} presents the architecture of the proposed SecLoop. Section \ref{3SA-GRPO} describes our proposed SA-GRPO algorithm. Section \ref{4exp} shows the experimental settings and discusses results compared with different baselines. Section \ref{5dis} gives some insightful discussions. Finally, conclusions are drawn in Section \ref{6con}.

\section{Our Proposed SecLoop System}
\label{2system}

\begin{figure*}[!t]
\centering
\includegraphics[width=\textwidth,trim=5 80 10 50,clip]{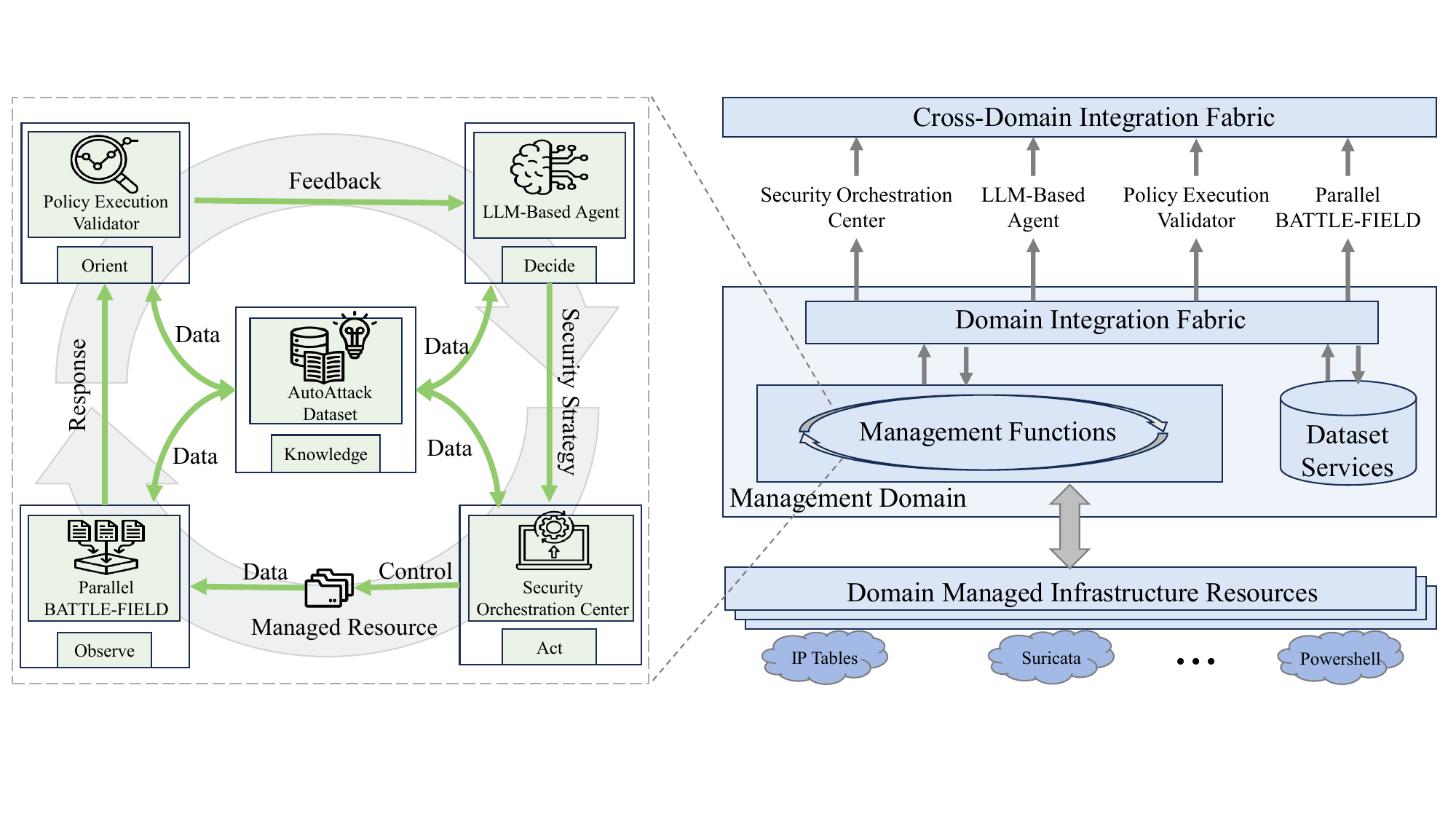}
\caption{Illustration of our framework mapped to ETSI ZSM. Data from the Parallel BATTLE-FIELD is fed into the policy execution validator. The analysis results from the validator are then delivered to the LLM-Based Agent, which generates strategic strategies. These strategies are executed by the SOC, and the final results are continuously monitored and recollected by the Parallel BATTLE-FIELD for subsequent cycles, constituting a closed-loop framework.}
\label{fig:7_mapping}
\end{figure*}

SecLoop is an advanced and automated security framework designed to address the dynamic and evolving security challenges in 6G ZTNs. As illustrated in Figure \ref{fig:2_system}, the system consists of three key modules: the parallel BATTLE-FIELD, the SOC, and LLM-guided strategy optimization. These components work together to provide a fully integrated, end-to-end solution for security strategy generation, execution, response, and continuous feedback. Figure \ref{fig:7_mapping} and Section \ref{sec:zsm_correspondence} describe the mapping relationship between the proposed SecLoop and the ZSM architecture. The details of the SOC and the overall system workflow are introduced in Section \ref{2sw}, while the specifics of the parallel BATTLE-FIELD are discussed in Section \ref{2bf}. The SA-GRPO algorithm, which drives the LLM-guided strategy optimization, is explained in detail in Section \ref{3SA-GRPO}. Additionally, Section \ref{2dataset} provides the construction process and details related to the training dataset. Through the interaction of these modules, SecLoop enhances its response to evolving threats while minimizing manual intervention.

\subsection{Correspondence with the ETSI ZSM Framework}
\label{sec:zsm_correspondence}

As illustrated in Figure \ref{fig:7_mapping},  we establish a correspondence between the SecLoop framework and the ETSI ZSM~\cite{etsi_zsm_002} framework. The right side of the figure depicts the ZSM framework, which is composed of managed infrastructure resources, management domain, dataset services, management functions, domain integration fabric, and cross-domain integration fabric. The core components of our SecLoop framework correspond to different logical functions within the ZSM's management functions, forming the loop illustrated in the left half of the figure. These components are: parallel BATTLE-FIELD, policy execution validator, LLM-based agent, and security orchestration center, which correspond respectively to the ZSM framework's domain data collection, domain analytics, domain intelligence, and domain orchestration. These components collaborate to form a complete observe, orient, decide, act (OODA) automation loop where the parallel BATTLE-FIELD collects data, the validator analyzes the situation, the agent generates policies, and the orchestration center executes them. This design clearly demonstrates a dynamic and adaptive zero-touch management process, validating the architectural feasibility and practical integration capability of the scheme.

\subsection{System Framework}
\label{2sw}

\begin{figure*}[!t]
\centering
\includegraphics[width=\textwidth,trim=122 155 122 50,clip]{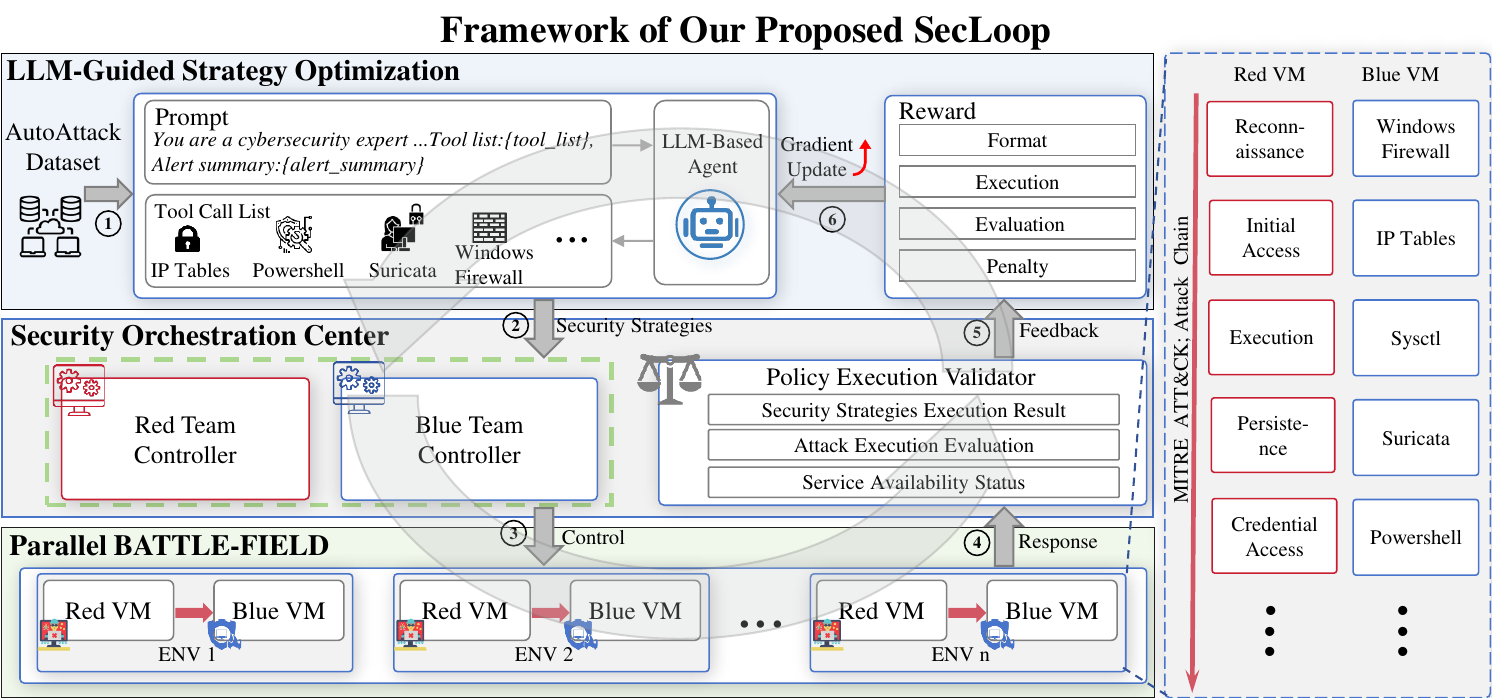}
\caption{Illustration of our proposed SecLoop. The system inputs are attack alerts from the AutoAttack dataset, which is fed into the LLM-based agent. LLM agent generates a group of strategies, which are then input into the SOC for execution. The red team controller and blue team controller automatically generate parallel BATTLE-FIELD environments to carry out the corresponding tool invocations. The response of BATTLE-FIELD is processed by the policy execution validator to generate feedback values, which are passed to the SA-GRPO reward function. SA-GRPO iteratively optimizes and updates the model parameters.}
\label{fig:2_system}
\end{figure*}

As illustrated in Figure \ref{fig:2_system}, the SecLoop framework begins by collecting attack alerts from the environments. These alerts are then processed and summarized into structured language descriptions, which serve as prompts for the LLM-based agent. Based on the input attack descriptions and the list of tools available in the current environments, the LLM agent generates a group of security strategies. Each strategy includes the tools to be invoked and their specific parameters in response to ongoing attacks. These strategies are fed into the red/blue team controller of the SOC, which automates the generation and parallel execution of BATTLE-FIELD. Each environment corresponds to a specific security strategy. Once the strategy is executed, the corresponding response from the environment is sent back to the policy execution validator in the SOC to generate feedback. The reward, based on strategies, feedback, and attack alerts, is used for model gradient updates. The LLM agent then refines the strategies iteratively to yield more optimal responses. We present the mathematical modeling of the entire SecLoop process below.

\textbf{1) Data Collection and Log Preprocessing:} First, for a set of attacks in the current environment, the n attack alerts collected by the IDS are represented as a set 
\begin{equation}
\mathcal{L} = \{L_1, \dots, L_i, \dots, L_n\},
\end{equation}
where each $L_i$ denotes an individual attack alert. Each $L_i \in \mathcal{L}$ is associated with specific attributes such as timestamp, attack type, and severity, forming a multi-dimensional feature space that can be used for further processing and analysis. These raw logs are typically noisy, repetitive, and heterogeneous in format. The summarizer module $Sr$ then extracts key event information, compresses redundant data, and outputs standardized m attack summaries, denoted as 
\begin{equation}
\mathcal{C} = Sr(\mathcal{L})=\{C_1, \dots, C_i , \dots, C_m\}, 
\end{equation}
where each $C_i$ represents a structured alert summary.

\textbf{2) Security Strategy Generation:} The summarized threat information $\mathcal{S}$, along with the current list of available tools 
\begin{equation}
\mathcal{T}_A = \{T_1, T_2, \dots, T_k\}
\end{equation}
and the environment description $\mathcal{E}$, are then formatted as prompt templates and input into the LLM based on Qwen2.5-7B-Instruct with fine-tuning. The concatenated input prompt is represented as
\begin{equation}
\mathbf{P} = [\mathcal{C}, \mathcal{T}_A, \mathcal{E}]. 
\end{equation}
The LLM processes the input prompt $\mathbf{P}$ through a function {\text{LLM}}, which generates a group of security strategies 
\begin{equation}
\mathcal{S} = {\text{LLM}}(\mathbf{P})= \{S_1, S_2, \dots, S_g\}. 
\end{equation}
Here, each strategy $S_i$ is a vector that includes the tool call list $\mathcal{T}_C^i$ to be invoked along with the associated parameters for mitigating the detected attacks.
\begin{equation}
\mathcal{T}_C^i = \{t_1, t_2, \dots, t_l\}, \quad \mathcal{T}_C^i \subseteq S_i, 
\end{equation}
where $t_j$ represents each tool involved in the strategy $S_i$ and $l$ is the number of tools in the tool call list $\mathcal{T}_C^i$.

\textbf{3) Simulation Verification and Multi-Dimensional Evaluation:} The generated security strategies $S_i$ are fed into the SOC via API interfaces for processing by the blue team controller (BTC) and red team controller (RTC). These controllers automate the generation of a set of parallel BATTLE-FIELD environments using infrastructure-as-code (IaC) templates defined by Vagrant, each consisting of virtual machines (VMs) representing the red and blue teams. Specifically, the red team environment is implemented using a red team virtual machine (RVM), while the blue team environment is represented by a blue team virtual machine (BVM).
\begin{equation}
\mathcal{E}_k = (E_{\text{RVM}}, E_{\text{BVM}})
\end{equation}
represents the k-th parallel execution environment, where $E_{\text{RVM}}$ and $E_{\text{BVM}}$ denote the red and blue team virtual machines, respectively. The BATTLE-FIELD environment is defined as:
\begin{equation}
E_{\text{BVM}} = \text{BTC}(S_k), \quad \text{for each} \quad S_k \in \mathcal{S}, 
\end{equation}
\begin{equation}
E_{\text{RVM}} = \text{RTC}(S_k), \quad \text{for each} \quad S_k \in \mathcal{S}, 
\end{equation}
where $S_k$ is the security strategy assigned to the environment $E_{\text{BVM}}$, and the execution is governed by the orchestrated interaction between the RVM and BVM.

After the execution of each strategy $S_i$, the corresponding response $RS_i =\{rs_{\text{exe}}(S_i), rs_{\text{attack}}(S_i), rs_{\text{service}}(S_i)\}$ from the environment is sent back to the policy execution validator in the SOC. This response consists of three key components: 1) security strategies execution result $rs_{\text{exe}}(S_i)$, 2) attack execution evaluation $rs_{\text{attack}}(S_i)$, and 3) service availability status $rs_{\text{service}}(S_i)$. These components provide essential feedback for assessing the effectiveness of the executed strategies.

As for the security strategy execution result, the success of each tool is represented as $f_{exe}(t_j) \in \{0, 1\}$, where the number 1 indicates successful execution and the number 0 indicates failure. The overall execution result $rs_{\text{exe}}(S_i)$ is the average success rate of all tools in $\mathcal{T}_C^i$:
\begin{equation}
rs_{\text{exe}}(S_i) = \frac{1}{l} \sum_{j=1}^{l} f_{exe}(t_j),\quad  t_j \in \mathcal{T}_C^i .
\end{equation}

To evaluate the success of the attack execution, we divide the attack process into multiple stages, each corresponding to a specific step in the attack chain. These stages are derived from the dataset. $\mathcal{P}_i = \{p_1, p_2, \dots, p_m\}$ represents the set of stages for the attack process associated with strategy $S_i$, and $p_j \in \mathcal{P}_i$ represents the j-th stage of the attack. The success of each attack stage is represented as $f_{exe}(p_j) \in \{0, 1\}$, where the number 1 indicates successful execution and the number 0 indicates failure. The overall execution result $rs_{\text{attack}}(S_i)$ is the average success rate of all attack stages in $\mathcal{P}_i$:
\begin{equation}
rs_{\text{attack}}(S_i) = \frac{1}{m} \sum_{j=1}^{m} f_{exe}(p_j),\quad p_j \in \mathcal{P}_i .
\end{equation}

The service availability status evaluates the impact of executing the strategy $S_i$ on the availability of network services, taking into account the effects in both the blue team and red team environments. After executing the strategy, the service availability status is represented as $rs_{\text{service}}(S_i) \in \{0, 1\}$, where the number 1 indicates that the service is available and unaffected by the attack, and the number 0 indicates that the service is disrupted or unavailable due to the attack.
\begin{equation}
rs_{\text{service}}(S_i) = f_{\text{service}}(E_{\text{BVM}}, E_{\text{RVM}}, S_i).
\end{equation}
Here, $f_{\text{service}}$ is a function that assesses the availability of services in the blue and red team environments after the execution of strategy $S_i$. The feedback is then represented as:
\begin{equation}
F_i =rs_{\text{exe}}(S_i)+ rs_{\text{attack}}(S_i)+ rs_{\text{service}}(S_i).
\end{equation}

\textbf{4) Feedback Optimization and Model Update:} The customized reward function in the SA-GRPO algorithm consists of four key components: format check, simulation execution, attack evaluation, and reasoning verification. The format check evaluates the format correctness of the security strategies $\mathcal{S}$, while simulation execution and attack evaluation reflect the degree to which the attack and defense strategies are executed in the environments, based on feedback $F$. Reasoning verification assesses how well the strategy aligns with the attack description $\mathcal{L}$, ensuring that the generated strategy matches the attack scenario. These components collectively determine the overall reward $Reward(S_i, F_i, \mathcal{L})$, which guides the optimization of strategies. The output of SA-GRPO is the gradient update $\Delta \theta$ for model parameter adjustment. Particularly, our proposed SA-GRPO algorithm, which integrates strategies comparison and a multi-dimensional reward function mechanism, enables the model to maintain high-quality responses while avoiding over-defense. This process can be triggered periodically or on demand, enabling the dynamic evolution of LLMs.

\subsection{Parallel BATTLE-FIELD}
\label{2bf}
Our BATTLE-FIELD is the core evaluation engine within the SecLoop framework, designed to construct a highly realistic, automated, and reusable cyber attack-defense simulation environment. The primary objective of this module is to provide a closed-loop validation platform for the security strategies under conditions closely resembling real-world cyberattacks. BATTLE-FIELD employs a red-blue confrontation mechanism, incorporating three key roles: attack simulation (Red Team), defence strategy (Blue Team), and evaluation. It simulates attack behaviors while executing defensive actions and quantitatively evaluates the effectiveness of the strategies.

To enable rapid deployment and environment reproducibility, BATTLE-FIELD leverages Vagrant to implement an Infrastructure-as-Code management mechanism. Vagrant allows users to define virtual network topologies through declarative configuration files, including host operating system types, IP address allocations, service configurations, and software installation scripts. With this capability, BATTLE-FIELD can automatically clone a complete experimental environment before each evaluation task begins, ensuring that comparisons between different strategies are conducted under identical or similar conditions. All virtual machine instances run on the VMware platform, utilizing its efficient resource scheduling and network isolation mechanisms to guarantee the independence and security of each experimental unit.

\textbf{1) Attack Simulation:} In terms of attack simulation, BATTLE-FIELD utilizes a custom-developed, lightweight platform as the red team's automated attack engine. This engine is built upon the MITRE ATT\&CK framework and is capable of automatically generating multi-stage attack chains based on predefined tactics and techniques. By loading different plugins and parameterized action profiles, the system can simulate typical attack behaviors such as initial access, privilege escalation, persistence, lateral movement, and data exfiltration. Moreover, our engine supports the development of dynamic and randomized attack modules. This allows researchers to design specific attack paths tailored to real-world scenarios, thereby enhancing the diversity and realism of the simulations and mitigating the risk of overfitting to a static attack configuration.

\textbf{2) Defense Strategy:} During a complete evaluation cycle, after the LLM-Guided Strategy Optimization submits the generated security policies to the SOC, the blue team controller first applies the corresponding mitigation measures on the blue team hosts located on BATTLE-FIELD based on the policy content, such as closing ports, updating firewall rules, and isolating suspicious hosts. Subsequently, the red team controller launches the predefined attack sequence using the automated attack engine and records key events during the attack process, such as successful compromise points and blocked attack locations. Meanwhile, the blue team controller monitors the system availability of the blue team hosts to comprehensively assess both the defensive effectiveness and side effects of the strategy.

\textbf{3) Evaluation:} After completion, the policy execution validator will collect all the report information, including the blue team's security strategies execution result, the red team's attack execution evaluation, and the blue team's service availability status, and then parse this report information to calculate the corresponding values and feed them back to the reward, which serves as the signal source for the execution reward and evaluation reward of the SA-GRPO algorithm, promoting the continuous improvement of the strategy model.

\subsection{Dataset Construction}
\label{2dataset}
We construct the AutoAttack dataset, which combines attack alerts with required automated environment configurations. This dataset is primarily designed for automating the replay of attack processes for BATTLE-FIELD, rather than for supervised model training with labeled data. The proposed AutoAttack integrates alert logs and real traffic, encompassing the automated replay of over 20 attack scenarios of 11 ATT\&CK processes, such as zero-day exploits. Our AutoAttack provides the necessary attack information inputs for LLMs. 

\textbf{1) Environment configuration for batch red-blue virtual machine pairing implemented by Vagrant:} The Vagrant tool is employed to create and configure the red and blue team virtual machines, allowing reproducible deployment of attack environments. All Vagrant-managed environments are defined and deployed through code, aligning with an Infrastructure-as-Code paradigm using provisioning scripts. This design encapsulates virtual machine configuration, network topology construction, and attack chain initialization into version-controllable artifacts such as shell and Python scripts. By avoiding manual configurations through GUI tools, this code-centric approach mitigates the risk of environment drift and enables reproducible and parameterized scenario generation.  
 
\textbf{2) Attack modules and MITRE ATT\&CK tactical mapping:} The MITRE ATT\&CK framework offers a structured taxonomy of adversarial tactics and techniques derived from real-world observations, serving as a standardized reference for describing attacker behaviors and organizing threat intelligence. Our approach integrates automated simulation of real-world cybersecurity incidents with synthetic network security event generation to map attack modules to MITRE ATT\&CK tactics. Automated simulations replicate complex attack behaviors, including nmap scanning, password cracking via SMB protocols. Synthetic traffic generation models browser-based XSS attacks and non-standard command-and-control (C2) patterns. This dual methodology ensures comprehensive alignment between attack modules and MITRE ATT\&CK tactics, as detailed in Table~\ref{tab:MITRE ATTCK}.

To address zero-day threats lacking public exploits, we developed a specialized synthetic event synthesis module. By analyzing technical disclosures and historical CVE data (e.g., CVE-2024-2961, CVE-2024-23897), we reconstruct high-impact vulnerability exploitation scenarios. Synthetic alerts emulate behavioral signatures of middleware services common to industrial and communication networks while evading rule-based detection. This enables evaluation of the SOAR strategy generalization against emerging, unseen threats.

\textbf{3) Automated Attack Orchestration:} We have developed a custom, lightweight automated attack engine for adversary emulation, built upon the MITRE ATT\&CK framework. Governed by the red team controller, this engine automatically orchestrates multi-stage attack chains by loading various plugins and parameterized action profiles. It simulates a wide range of APT behaviors and executes them in the red-team VM to conduct realistic, threat-based stress tests on the blue-team VM. A key feature is its support for dynamic and randomized attack modules, which enhances simulation realism and mitigates the risk of overfitting to static configurations. This system effectively bridges theoretical ATT\&CK structures with scalable and dynamic adversary emulation, overcoming the limitations of traditional human-computer interaction.

\begin{table}
\centering
\caption{Attack Stages and MITRE ATT\&CK Tactical Mapping}
\label{tab:MITRE ATTCK}
\begin{tabular}{c|l}
\hline
\multicolumn{1}{c|}{\textbf{\begin{tabular}[c]{@{}c@{}}MITRE ATT\&CK \\ Tactical Stages\end{tabular}}} & 
\multicolumn{1}{c}{\textbf{\begin{tabular}[c]{@{}c@{}}Attack Means\end{tabular}}}  \\
\hline
\textit{\begin{tabular}[c]{@{}l@{}}Reconnaissance\end{tabular}} & \textit{\begin{tabular}[c]{@{}l@{}}Port scanning through fscan/nmap\end{tabular}} \\
\hline
\textit{\begin{tabular}[c]{@{}l@{}}Initial Access\end{tabular}} & \textit{\begin{tabular}[c]{@{}l@{}}Redis unauthorized vulnerability written \\to webshell Trojan, CVE-2025-29927\end{tabular}} \\
\hline
\textit{\begin{tabular}[c]{@{}l@{}}Execution\end{tabular}} & \textit{\begin{tabular}[c]{@{}l@{}}File upload and write to webshell Trojan, \\Deserialization attack, \\CVE-2024-23897, CVE-2025-24813\end{tabular}} \\
\hline
\textit{\begin{tabular}[c]{@{}l@{}}Persistence\end{tabular}} & \textit{\begin{tabular}[c]{@{}l@{}}Preliminary webshell Trojan Writing \\Complete Logic for webshell Control Trojan, \\CVE-2024-2961\end{tabular}} \\
\hline
\textit{\begin{tabular}[c]{@{}l@{}}Credential Access\end{tabular}} & \textit{\begin{tabular}[c]{@{}l@{}}XSS (Cross Site Script), \\Cross Site Request Forgery, etc.\end{tabular}} \\
\hline
\textit{\begin{tabular}[c]{@{}l@{}}Discovery\end{tabular}} & \textit{\begin{tabular}[c]{@{}l@{}}SQL injection, LFI/RFI, SSRF,\\ CVE-2024-23897, CVE-2025-30208\end{tabular}} \\
\hline
\textit{\begin{tabular}[c]{@{}l@{}}Lateral Movement\end{tabular}} & \textit{\begin{tabular}[c]{@{}l@{}}Blasting Windows account passwords \\by CrackMapExec through SMB protocol\end{tabular}} \\
\hline
\textit{\begin{tabular}[c]{@{}l@{}}Collection\end{tabular}} & \textit{\begin{tabular}[c]{@{}l@{}}C2 Trojan collects information from \\Blue Team's target drone\end{tabular}} \\
\hline
\textit{\begin{tabular}[c]{@{}l@{}}Command and Control\end{tabular}} & \textit{\begin{tabular}[c]{@{}l@{}}C2 Trojan was activated and went live \\on the blue team's target machine\end{tabular}} \\
\hline
\textit{\begin{tabular}[c]{@{}l@{}}Exfiltration\end{tabular}} & \textit{\begin{tabular}[c]{@{}l@{}}Transmitting sensitive information of \\the target drone through C2\end{tabular}} \\
\hline
\textit{\begin{tabular}[c]{@{}l@{}}Impact\end{tabular}} & \textit{\begin{tabular}[c]{@{}l@{}}Man in the middle attacks in the form of \\DNS hijacking, DoS/DDoS Attacks\end{tabular}}\\
\hline
\end{tabular}
\end{table}

To reflect the operational characteristics of communication network environments, we apply a filtering mechanism to exclude techniques with limited relevance to network traffic analysis, such as Resource Development and Privilege Escalation, and instead focus on failure-prone defense scenarios frequently encountered in the communication field. 

\section{SA-GRPO Algorithm Design}
\label{3SA-GRPO}

\begin{figure*}[!t]
\centering
\includegraphics[width=\textwidth]{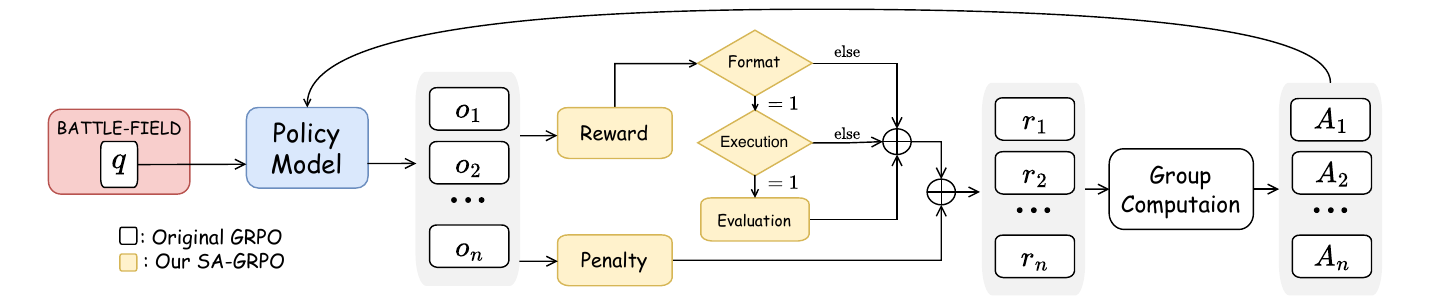}
\caption{Illustration of our proposed SA-GRPO algorithm. A group of outputs is sampled from the policy model, each assigned a reward computed with four customized reward functions. The group relative advantage is estimated for each output, and the policy model is updated by maximizing the objective function.}
\label{fig:3_sagrpo}
\end{figure*}
\subsection{Workflow of SA-GRPO}

Traditional supervised fine-tuning (SFT) methods rely heavily on a high-quality labeled dataset and bring significant GPU memory consumption. In contrast, the GRPO algorithm leverages the comparative results of a group of outputs, enabling more efficient learning from diverse scenarios without requiring extensive labeled datasets. Based on GRPO, we propose SA-GRPO, a security-aware group relative policy optimization algorithm tailored specifically for security automation of SecLoop. As shown in Figure \ref{fig:3_sagrpo}, SA-GRPO updates the policy model by maximizing the SA-GRPO objective, which is computed using an estimator of the advantage based on the relative rewards of outputs within each group. Detailed procedures for our proposed SA-GRPO are outlined below.

\begin{algorithm}[htbp]
\caption{Our Proposed SA-GRPO}\label{alg:SA-GRPO}
\begin{algorithmic}
\STATE \textbf{Input:} initial Policy Model $\pi_\theta$, Reward Model $R$, Task Prompts $\mathcal{D}$,
Hyperparameters $\epsilon_{\text{low}}, \epsilon_{\text{high}}$
    \FOR{step = 1, \ldots, $M$}  
        \STATE $D_b \sim \mathcal{D}$
        \STATE $\pi_{\theta_{\text{old}}} \gets \pi_\theta$
        \FOR {each question $q \in \mathcal{D}_b$}
            \STATE $\{o_i\}_{i=1}^G \sim \pi_{\theta_{\text{old}}}(\cdot | q)$ 
        \ENDFOR
        \FOR{each $o_i$}
            \STATE $\left\{ r_i \right\}_{i=1}^{G} \gets \left\{ R(q, o_i) \right\}_{i=1}^{G}$
            \STATE $\hat{A}_{i,t} \gets {r_i-\mathrm{mean}(\{r_i\}_{i=1}^{G})}$
        \ENDFOR
        \FOR{SA-GRPO iteration $= 1, \ldots, \mu$}
            \STATE $\pi_{\theta} \leftarrow \mathop{\arg\max}\limits_{\theta} \mathcal{J}_{\mathrm{SA-GRPO}}(\theta)$
        \ENDFOR
    \ENDFOR
    \STATE \textbf{Output:} $\pi_\theta$
\end{algorithmic}
\end{algorithm}

SA-GRPO introduces a policy model $\pi_\theta$, a reward model, a group of task prompts $\mathcal{D}$ and hyperparameters such as $\epsilon_{\text{low}}, \epsilon_{\text{high}}$. For each subsequent step, SA-GPRO samples a batch $\mathcal{D}_b$ from $\mathcal{D} $ and updates the former policy model. Then it samples a group of outputs $\{o_i\}_{i=1}^{G}$ for each question $q$ when using old policy $\pi_{\theta_\mathrm{old}}$ and optimizes the policy via the following objective:
 \begin{align}
 \mathcal{J}_{\mathrm{SA-GRPO}}(\theta) &= 
    \mathbb{E}_{(q,a)\sim\mathcal{D},\, \{o_i\}_{i=1}^G \sim \pi_{\theta_\mathrm{old}}(\cdot|q)} \notag
    \\
    \qquad\Biggl[ 
    \frac{1}{G} 
    \sum_{i=1}^{G} \sum_{t=1}^{|o_i|} 
    &\min \Bigl( \gamma_{i,t}(\theta) \hat{A}_{i,t},\mathrm{clip}_{i,t}(\theta) \hat{A}_{i,t} \Bigr)
     \Biggr],  \label{eq:sa-grpo-objective}
    \end{align}
where $(q,a)$ is a question-answer pair from the data distribution $\mathcal{D}$, and $\{o_i\}_{i=1}^{G}$ are G outputs sampling for each question-answer pair $(q,a)$; $\mathrm{clip}_{i,t}(\theta)$ is a clip operation for importance sampling ratio $ \gamma_{i,t}(\theta)$, where
\begin{equation}
     \mathrm{clip}_{i,t}(\theta)= \mathrm{clip}\bigl(\gamma_{i,t}(\theta),\, 1 - \varepsilon_\mathrm{low},\, 1 + \varepsilon_\mathrm{high}\bigr),
\end{equation}
where $\varepsilon_\mathrm{low}$ and $\varepsilon_\mathrm{high}$ are respectively hyperparameters within the lower and upper bound of the clipping range of the importance sampling ratio $ \gamma_{i,t}(\theta)$ for the $i$-th output at the $t$-th token,
where
\begin{equation}
    \gamma_{i,t}(\theta)=    \frac{\pi_\theta(o_{i,t} \mid {q, o_{i,<t}})}          {\pi_{\theta_\mathrm{old}}(o_{i,t} \mid {q, o_{i,<t}})},
\end{equation}
where $\pi_\theta(o_{i,t}\mid{q, o_{i,<t}})$ is the probability of $i$-th object at $t$-th token under the conditions of question $q$ and object $o_i$ when using updated policy $\pi_\theta$ while $\pi_{\theta_\mathrm{old}}(o_{i,t} \mid{q, o_{i,<t}})$ is the probability when using old policy $\pi_{\theta_\mathrm{old}}$; SA-GRPO then computes $\hat{A}_{i,t}$ which is an estimator of the advantage at time step $t$ for each output $o_i$, where 
\begin{equation}
    \hat{A}_{i,t} = 
   {R({q,o_i})-\mathrm{mean}(\{R({q,o_i})\}_{i=1}^{G})}. \label{eq:advantage}
\end{equation}
Given the reward function $R$, $\hat{A}_{i,t}$ is calculated based on the relative rewards of the outputs inside each group. 
\subsection{Reward Modules Design} 

\noindent
\textbf{1) Format Reward:} The format reward evaluates the structural correctness of the outputs generated by the LLM, specifically assessing whether they adhere to the expected format (JSON). The score is assigned on a scale from 0 to 1, with a score of 1 indicating complete conformity to the predefined structure. Only outputs that achieve the full score of 1 are eligible to proceed to the subsequent reward module. Otherwise, the response advances to the final penalty stage. To automate format verification and ensure consistency, we employ regular expressions as the validation mechanism. The format reward function $R_\mathrm{format}$ takes outputs $o_i$ of LLM as inputs which is defined as:
\begin{equation}
    R_\mathrm{format}(o_i)=
    \begin{cases}
1, & \text{if } re(o_i) = true\\
0, & \text{if } re(o_i) = false
\end{cases},
\end{equation}  
where $re(o_i)$ is a function based on regular expressions that validates the format of $o_i$.

\noindent
\textbf{2) Execution Reward:} The execution reward assesses the ability of the LLM-generated instructions to execute correctly within our BATTLE-FIELD. The reward is conducted on a scale from 0 to 1, where a score of 1 is assigned only if the instruction is executed without error. Instructions that fail execution advance to the final penalty stage. To ensure robust execution verification and maintain consistency, the reward module is closely integrated with the BATTLE-FIELD. The execution reward function can be defined as $R_\mathrm{exec}$, which takes outputs from LLM as inputs $o_i$ where
\begin{equation}
    R_\mathrm{exec}(o_i)=
    \begin{cases}
1, & \text{if } E(o_i) = true\\
0, & \text{if } E(o_i) = false
\end{cases},
\end{equation}
where $E(o_i)$ is a function that verifies the executable instructions in $o_i$.

\noindent
\textbf{3) Evaluation Reward:} The Evaluation Reward measures the effectiveness of LLM-generated instructions in defending against and mitigating attacks within the BATTLE-FIELD. This reward is quantified on a scale from 0 to 1, where a score of 1 indicates the absence of warning alarms following the execution of the provided instructions. Lower scores are assigned based on the degree to which the instructions successfully enhance the environment's defensive and mitigation capabilities. Subsequently, the instructions proceed to the final penalty assessment stage. To ensure rigorous validation of execution outcomes, the reward module is tightly integrated with the BATTLE-FIELD platform. The evaluation reward function can be defined as $R_\mathrm{eva}$, which takes outputs from LLM as inputs $o_i$ where
\begin{equation}
    R_\mathrm{eva}(o_i)=W_\mathrm{BATTLE-FIELD}(o_i) \in [0,1],
\end{equation}
where $W_\mathrm{BATTLE-FIELD}(o_i)$ is a function based on BATTLE-FIELD that evaluates the
defensive and mitigation capabilities of instructions in $o_i$.

\noindent
\textbf{4) Penalty:} To prevent the excessive processing of instructions generated by the LLM, we introduce a penalty module designed to mitigate potential biases and enhance the overall robustness of our system. An automated LLM-based expert is employed to evaluate the generated instructions, ensuring their validity and appropriateness. This approach effectively avoids extreme or overly harsh operations, thereby maintaining balanced and reliable system performance. The penalty function can be defined as $ P(o_i)$ which takes outputs from LLM as inputs $o_i$ where
\begin{equation}
    P(o_i)= 1- W_\mathrm{LLM}(o_i), W_\mathrm{LLM} (o_i)\in [0,1],
\end{equation}
where $W_\mathrm{LLM}(o_i)$ is a function based on an LLM-based expert that evaluates the rationality
 of instructions in $o_i$.
\section{Experiments}
\label{4exp}
\subsection{Experimental Setup}

\textbf{1) Configuration of LLM Training:} The experimental setup of the SecLoop framework is designed to support distributed training, high-fidelity simulation, and efficient policy validation. The training node is deployed on a CentOS Stream 8 server equipped with 8 NVIDIA A800 GPUs, 256GB DDR4 memory, and 10TB NVMe SSD storage. This node runs PyTorch 2.3 and HuggingFace Transformers, with model architecture based on Qwen2.5-7B-Instruct. Multi-GPU communication is accelerated via NCCL libraries, while Slurm manages job scheduling.

\textbf{2) Configuration of LLM Evaluation:} The evaluation node operates on an Ubuntu 22.04 server with 1 NVIDIA A100 GPU (80GB HBM2e), 64GB DDR4 memory, and 1TB NVMe SSD. This node executes security policies generated by the training node, hosts a lightweight communication middleware.

\textbf{3) Configuration of BATTLE-FIELD:} The BATTLE-FIELD simulation environment is hosted on a Windows 11 Pro machine (Intel Core i7-14700F, 64GB DDR4 memory) integrated with Vagrant 2.4.3 and VMware Workstation Pro 17.5. This setup enables dynamic cloning of red-team and blue-team virtual machines through Infrastructure-as-Code templates.

\textbf{4) Configuration of Policy Execution Validator:} The red team controller and the blue team controller are deployed on separate Ubuntu 22.04 servers, each of which has 16GB of memory and a 512GB NVMe SSD. The red team controller coordinates our self-developed attack agent to execute the specified attack strategy. The blue team controller performs defensive actions. Their execution results will be fed back to the policy execution validator. The policy execution validator resides in a WSL2 subsystem on the Windows 11 host, running Ubuntu 22.04 with 64GB of memory.

\textbf{5) Configuration of Real-World Test:} For real-world deployment testing, we evaluated SecLoop on a set of embedded edge devices, including four Jetson AGX Orin 64GB and four Jetson ORIN NX 16GB. Each Jetson AGX Orin 64GB features a 2048-core NVIDIA Ampere GPU with 64 Tensor Cores and a 12-core Arm Cortex-A78AE CPU, while the Jetson ORIN NX features a 1024-core GPU and a 6-core CPU. In addition, we included a Windows 11 laptop with 16GB RAM and an Ubuntu 22.04 desktop machine with 64GB RAM in our evaluation, further demonstrating the framework’s cross-platform compatibility and adaptability across heterogeneous hardware environments.

\subsection{Datasets}
The attack datasets CIC-IDS2017~\cite{rosay2022network}, CIC-IDS2018~\cite{leevy2020survey}, UNSW-NB15~\cite{moustafa2015unsw}, and CCDC-2018~\cite{cover1999elements}, which are widely used in the current field of network security, are respectively collected for network attack data in different scenarios. CIC-IDS2017 and CIC-IDS2018 mainly collect attack traffic in the conventional network environment, including various attack types such as DDoS, DoS, penetration attacks, brute-force cracking, and Web attacks; UNSW-NB15 covers nine types of modern cyber attacks, such as missed strike attacks, worm spread, and Shellcode injection. CCDC-2018 is based on a real offensive and defensive exercise environment and records multi-stage compound attacks such as phishing attacks, malware spread, and lateral penetration.

We also construct corresponding attack environments in BATTLE-FIELD based on the datasets, enabling faithful replay of attacks and accurate evaluation of generated outputs. In comparison with baseline reinforcement learning algorithms such as PPO and KTO, we incorporate Reinforcement Learning from Human Feedback (RLHF). In this approach, domain experts are engaged to annotate the training dataset with human preferences, which are subsequently leveraged to generate human-guided rewards during the training phase.

\begin{table*}
\centering
\caption{Comparisons of Security Automation Systems}
\label{tab:secloop}
\begin{tabular}{c|c|c|c|c|c}
\hline
\multicolumn{1}{c|}{\textbf{Scheme}} & \textbf{\begin{tabular}[c]{@{}c@{}}Strategy\\Generation\end{tabular}} & \textbf{\begin{tabular}[c]{@{}c@{}}Automated Security \\ Orchestration\end{tabular}} & {\textbf{Response}} & {\textbf{Feedback}} & \textbf{\begin{tabular}[c]{@{}c@{}}ATT\&CK Stages\end{tabular}} \\
\hline
\textit{\begin{tabular}[c]{@{}l@{}}TENNISON~\cite{fawcett2018tennison}\end{tabular}} & \checkmark & \checkmark & \checkmark & \checkmark & \textit{\begin{tabular}[c]{@{}c@{}}2, 3, 4, and 6\end{tabular}} \\
\hline
\textit{\begin{tabular}[c]{@{}c@{}}JESS~\cite{kalkan2018jess}\end{tabular}} & \checkmark & $\times$ & \checkmark & $\times$ & \textit{\begin{tabular}[c]{@{}c@{}}1, 2, and 13\end{tabular}} \\
\hline
\textit{\begin{tabular}[c]{@{}c@{}}Virtual IoT HoneyNets~\cite{zarca2020virtual}\end{tabular}} & \checkmark & \checkmark & \checkmark & $\times$ & \textit{\begin{tabular}[c]{@{}c@{}}1, 2, 3, 4, 5, 6, and 13\end{tabular}} \\
\hline
\textit{\begin{tabular}[c]{@{}c@{}}OntoCSD~\cite{wu2024ontocsd}\end{tabular}} & \checkmark & \checkmark & \checkmark & $\times$ & \textit{\begin{tabular}[c]{@{}c@{}}1, 2, 4, 5, 6, 9, and 13\end{tabular}} \\
\hline
\textit{\begin{tabular}[c]{@{}c@{}}APIRO~\cite{sworna2023apiro}\end{tabular}} & \checkmark & $\times$ & \checkmark & $\times$ & \textit{\begin{tabular}[c]{@{}c@{}}1, 2, 3, 4, 5, 9, and 13\end{tabular}} \\
\hline
\textit{\begin{tabular}[c]{@{}c@{}} AG-AEGM~\cite{liu2024generic}\end{tabular}} & \checkmark & \checkmark & \checkmark & $\times$ & \textit{\begin{tabular}[c]{@{}c@{}}1, 2, 3, 4, 5, 6, 7, 9, 10, and 13\end{tabular}} \\
\hline
\textit{\begin{tabular}[c]{@{}c@{}}SSAE-SVM~\cite{long2022hybrid}\end{tabular}} & \checkmark & $\times$ & $\times$ & $\times$ & \textit{\begin{tabular}[c]{@{}c@{}}13\end{tabular}} \\
\hline
\textit{\begin{tabular}[c]{@{}l@{}}IoT-DPS~\cite{otoum2025llm}\end{tabular}} & \checkmark & \checkmark & \checkmark & $\times$ & \textit{\begin{tabular}[c]{@{}c@{}}1, 7, 8, 11, and 13\end{tabular}} \\
\hline
\textit{\begin{tabular}[c]{@{}l@{}}RAG-IR~\cite{tellache2025advancing}\end{tabular}} & \checkmark & \checkmark & \checkmark & $\times$ & \textit{\begin{tabular}[c]{@{}c@{}}2, 3, 5, 6, 7, 8, 9, 10, 11, and 12\end{tabular}} \\
\hline
\textit{\begin{tabular}[c]{@{}l@{}}IRCopilot~\cite{lin2025ircopilot}\end{tabular}} & \checkmark & \checkmark & \checkmark & \checkmark & \textit{\begin{tabular}[c]{@{}c@{}}3, 4, 5, and 6\end{tabular}} \\
\hline
\textit{\begin{tabular}[c]{@{}l@{}}ReAct-LLM~\cite{baral2025autonomous}\end{tabular}} & \checkmark & \checkmark & \checkmark & $\times$ & \textit{\begin{tabular}[c]{@{}c@{}}1, 2, 7, and 8\end{tabular}} \\
\hline
\textbf{\textit{SecLoop(Ours)}} & \checkmark & \checkmark & \checkmark & \checkmark & \textit{\begin{tabular}[c]{@{}c@{}}1, 2, 3, 4, 7, 8, 9, 10, 11, 12, and 13\end{tabular}} \\
\hline
\end{tabular}

\begin{tablenotes}
        \footnotesize
        \item The numbers 1-13 correspond to the ATT\&CK tactical stages: Reconnaissance, Initial Access, Execution, Persistence, Privilege Escalation, Defense Evasion, Credential Access, Discovery, Lateral Movement, Collection, Command and Control, Exfiltration, and Impact.
\end{tablenotes}

\end{table*}

\begin{figure*}[!t]
\centering
\subfloat[]{\includegraphics[width=2.2in]{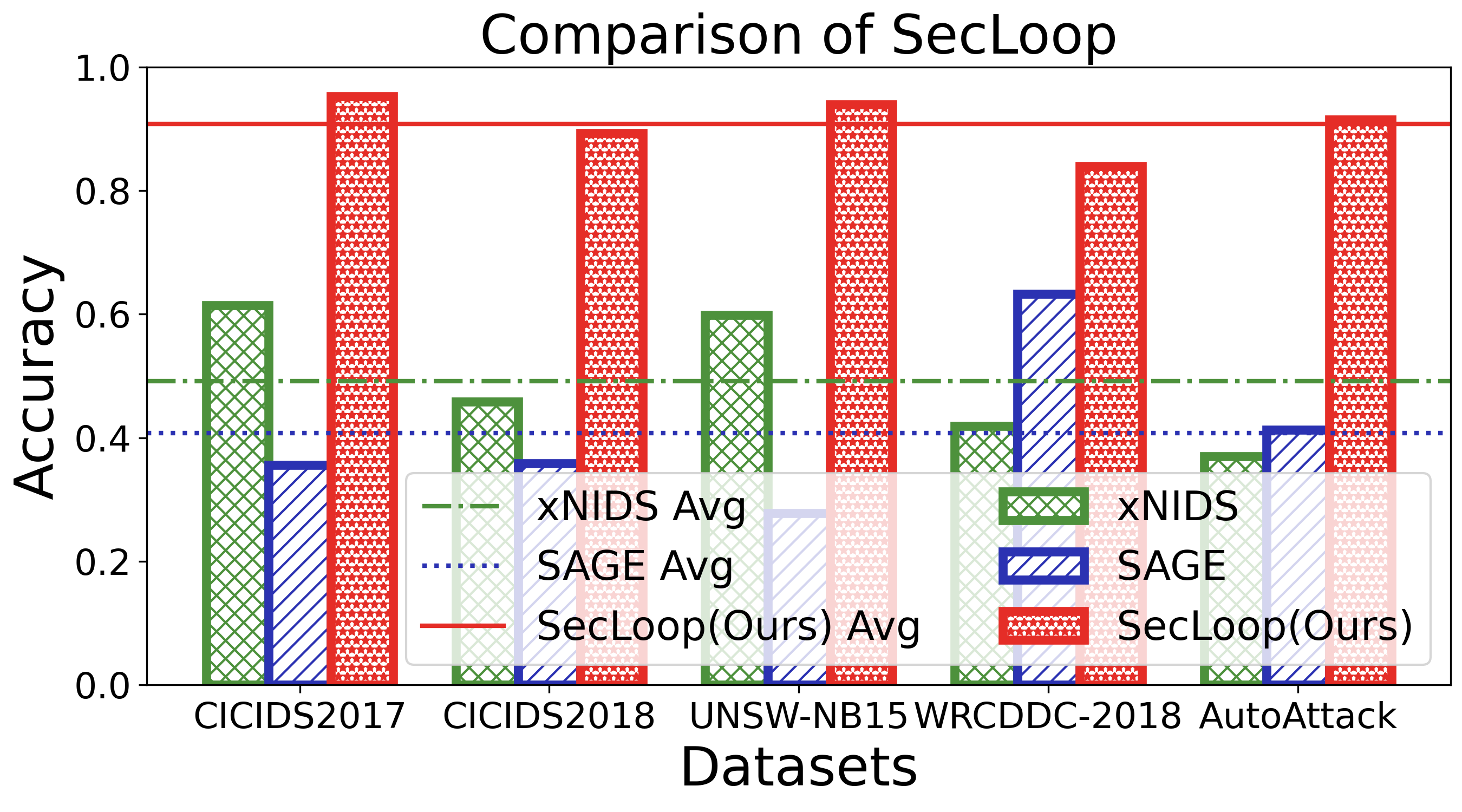}
\label{fig:4_first_case}}
\hfil
\subfloat[]{\includegraphics[width=2.2in]{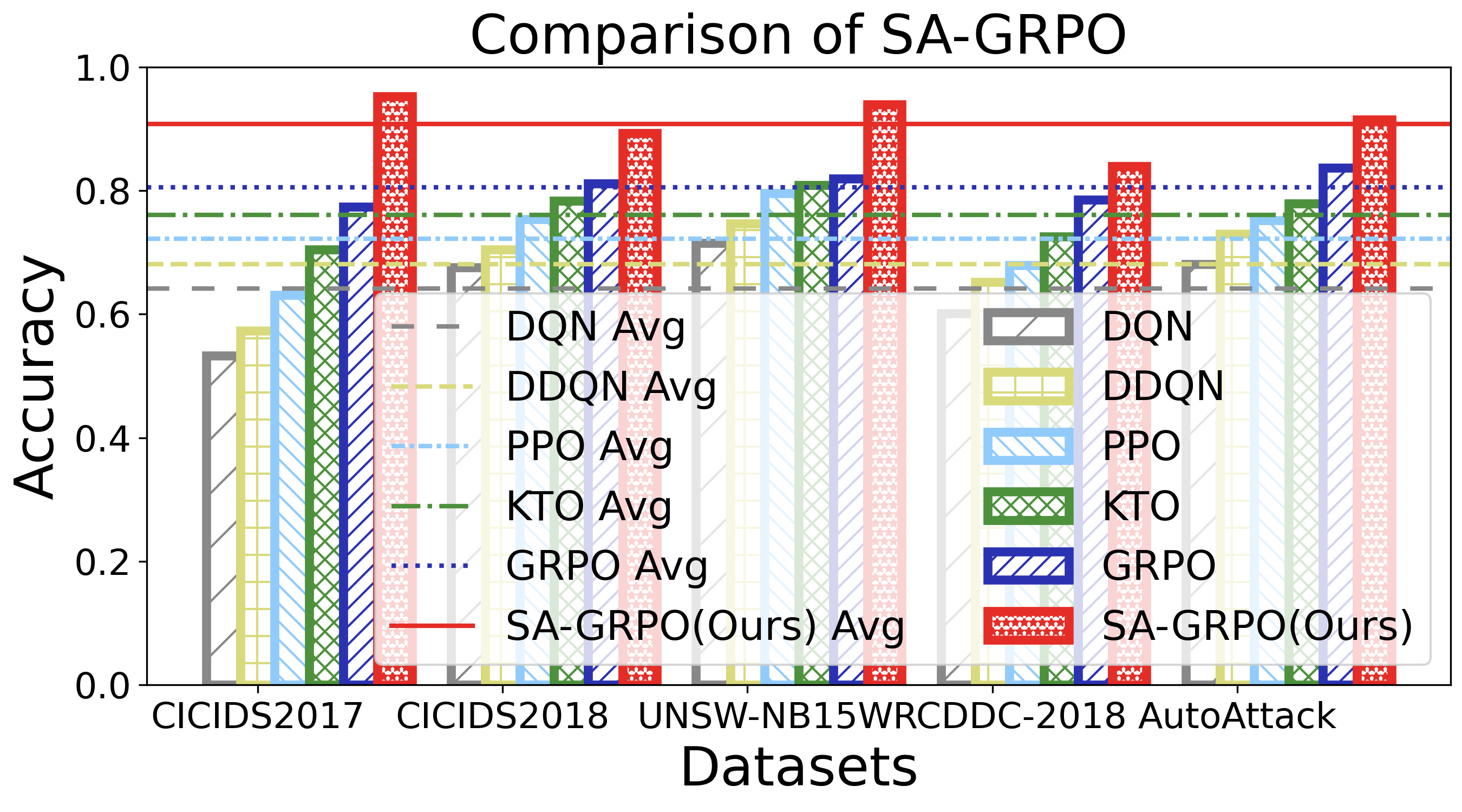}
\label{fig:4_second_case}}
\hfil
\subfloat[]{\includegraphics[width=2.2in]{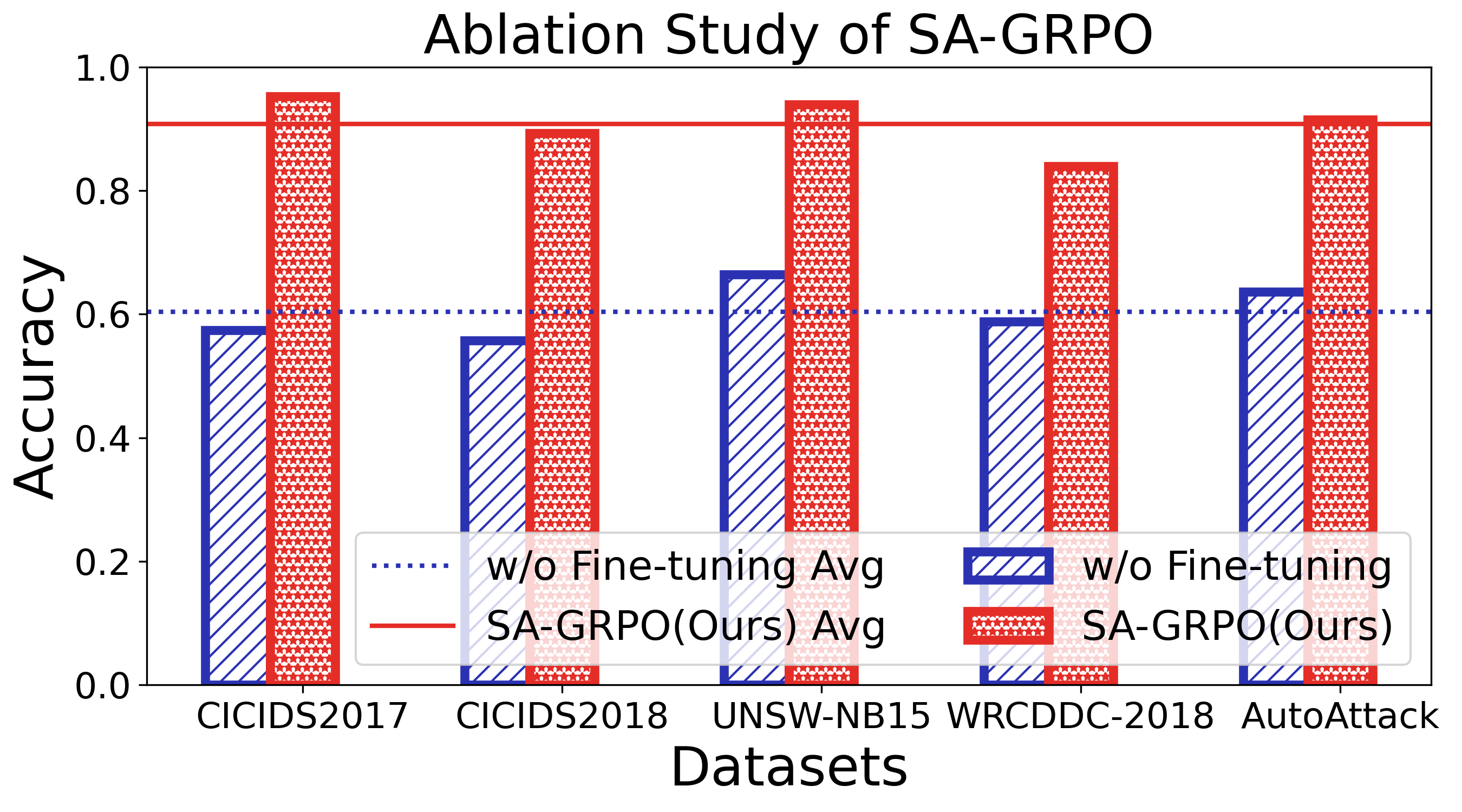}
\label{fig:4_third_case}}
\hfil
\subfloat[]{\includegraphics[width=2.2in]{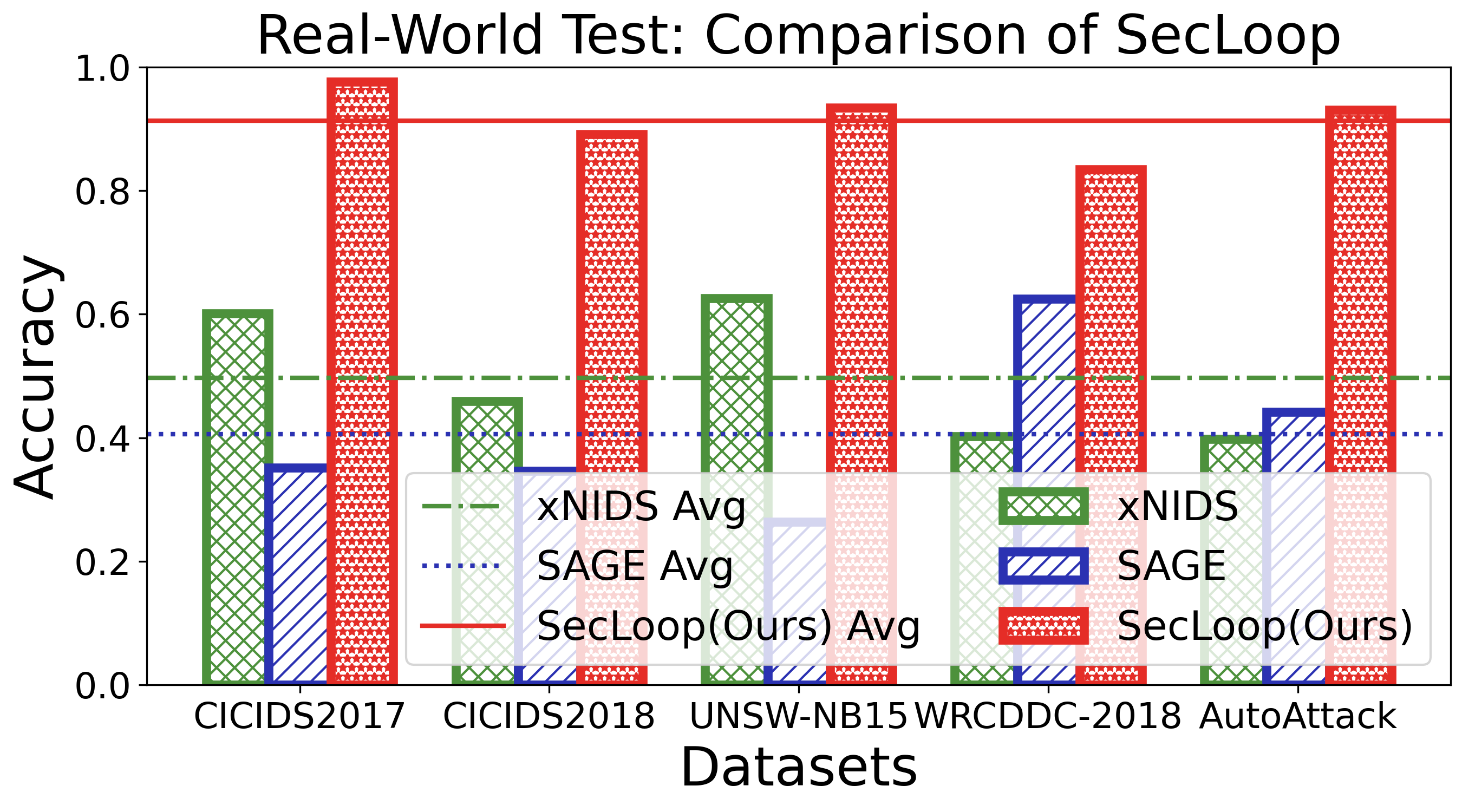}
\label{fig:4_fourth_case}}
\hfil
\subfloat[]{\includegraphics[width=2.2in]{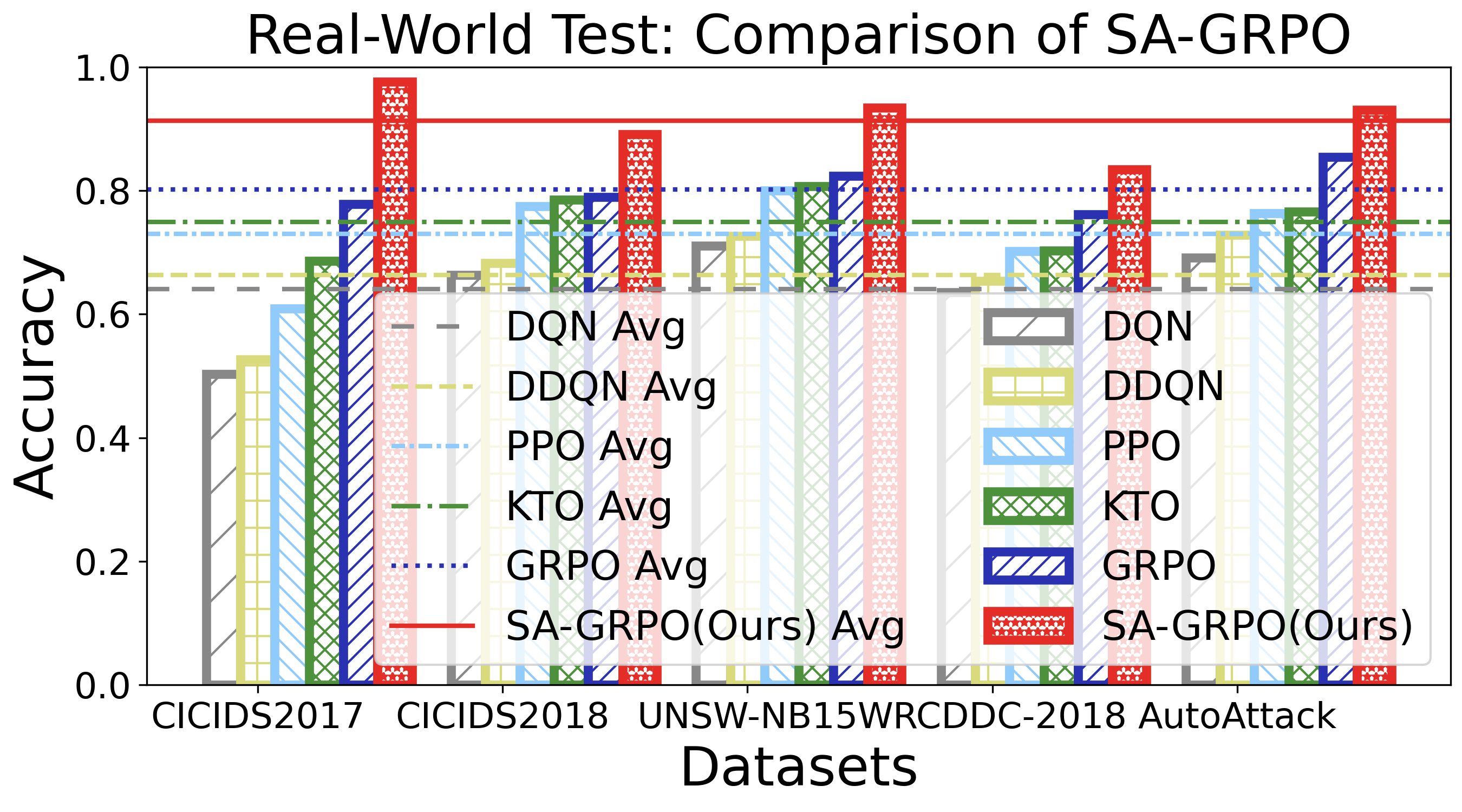}
\label{fig:4_fifth_case}}
\hfil
\subfloat[]{\includegraphics[width=2.2in]{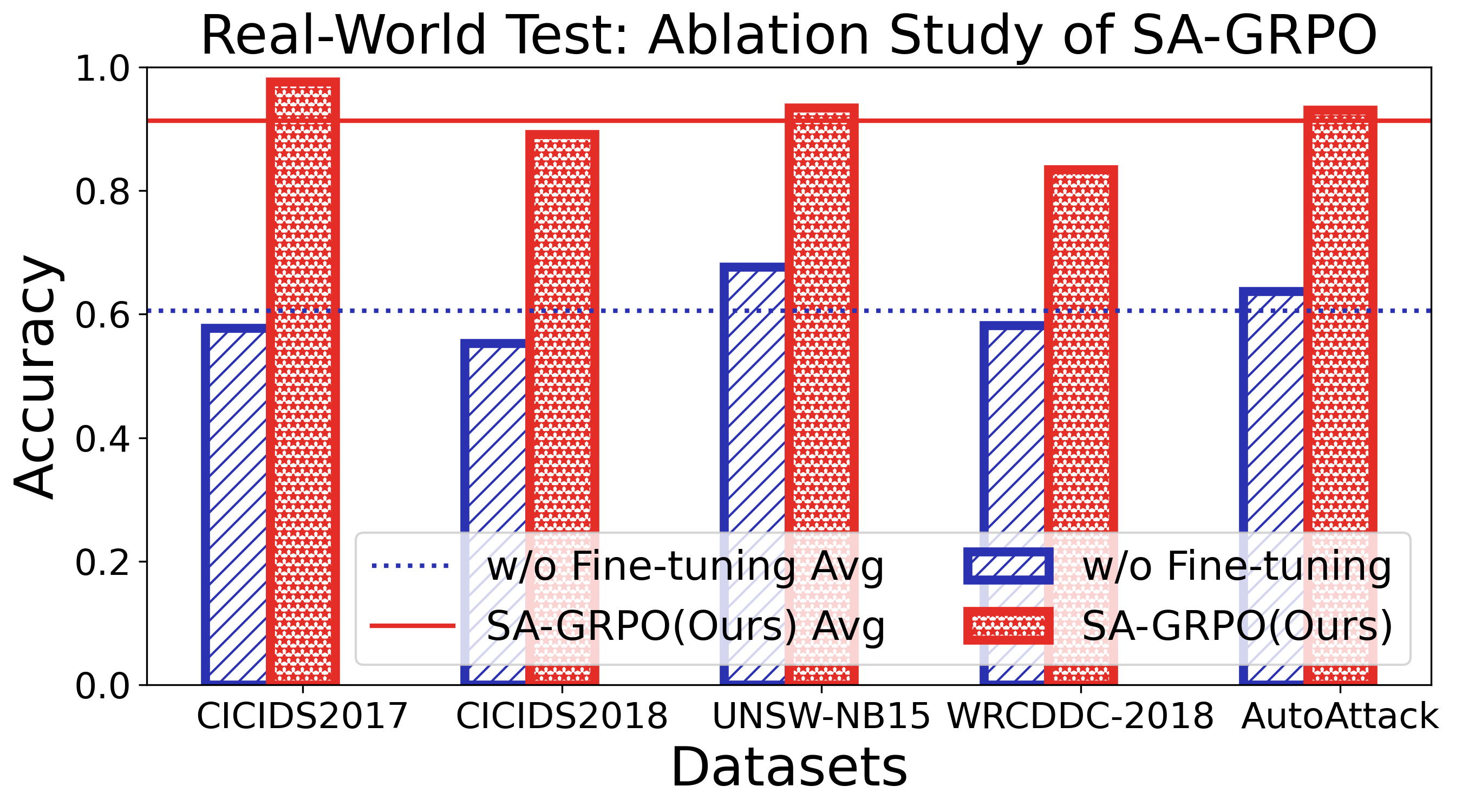}
\label{fig:4_sixth_case}}
\caption{Performance comparisons and ablation studies of our proposed SecLoop and SA-GRPO. (a), (b), and (c) compare the accuracy of SecLoop and SA-GRPO against multiple baselines on five benchmarks. (d), (e), and (f) evaluate the accuracy of SecLoop and SA-GRPO in the real-world testbed. }
\label{fig:4}
\end{figure*}

\subsection{Baselines}
We compare the function of our proposed SecLoop with various baselines, including TENNISON~\cite{fawcett2018tennison}, JESS~\cite{kalkan2018jess}, Virtual IoT HoneyNets~\cite{zarca2020virtual}, OntoCSD~\cite{wu2024ontocsd}, APIRO~\cite{sworna2023apiro}, AG-AEGM~\cite{liu2024generic}, SSAE-SVM~\cite{long2022hybrid}, IoT-DPS~\cite{otoum2025llm}, RAG-IR~\cite{tellache2025advancing}, IRCopilot~\cite{lin2025ircopilot}, and ReAct-LLM~\cite{baral2025autonomous}. TENNISON employs adaptive distributed strategies against DoS/DDoS/scanning via multi-level detection and orchestration. JESS uses three-stage mitigation (Nominal-Preparatory-Active) with joint entropy analysis for DNS/NTP attacks. Virtual IoT HoneyNets deploy policy-driven honeypots with real-time monitoring for IoT botnets/zero-days. OntoCSD combines CBR reasoning with OWL/SWRL threat modeling for credential attacks/phishing. APIRO integrates NLP-based API recommendations with heterogeneous toolsets for data breach mitigation. AG-AEGM evaluates defenses through attack graph analysis and evolutionary game theory. SSAE-SVM employs entropy analysis plus deep learning (SSAE/SVM) for hybrid attack detection. IoT-DPS utilizes lightweight LLMs to enable real-time detection and automatic response to abnormal traffic and attack behaviors in IoT networks. RAG-IR dynamically integrates LLMs with CTI to achieve automated, context-aware security alert enhancement and response strategy generation. IRCopilot applies LLMs to the automation of event response throughout the entire lifecycle. React-LLM is an autonomous network event response system based on the ReAct (Reasoning + Acting) framework and LLMs. Our system uniquely automates strategy validation through integrated response-feedback loops in security orchestration.

We compare the performance of our proposed SecLoop with state-of-the-art baselines, including xNIDS~\cite{wei2023xnids} and SAGE~\cite{nadeem2021enabling}. xNIDS transfers the results of deep learning-based intrusion detection systems to actionable responses. SAGE utilizes an unsupervised S-PDFA model to compress intrusion alerts into attack graphs for strategy generation. Different from baselines, SecLoop can generate defense strategies that enable automation tool execution in diverse environments.

The proposed SA-GRPO is compared with several reinforcement learning approaches, including GRPO~\cite{shao2024deepseekmath}, KTO~\cite{ethayarajh2024model}, DQN~\cite{mnih2013playing}, DDQN~\cite{van2016deep}, and PPO~\cite{schulman2017proximal}. PPO uses a clipped surrogate objective to constrain policy updates within a proximal region of the previous policy. KTO maximizes the utility of generations instead of maximizing the log-likelihood of preferences. DQN learns off-policy Q-values with replay and a target network for discrete actions, but tends to overestimate. DDQN decouples selection and evaluation to reduce bias and stabilize training. Unlike PPO and KTO, GRPO adopts a group-relative advantage estimation approach. Specifically, our proposed SA-GRPO algorithm is a redesign of GRPO tailored for SecLoop, which eliminates the division by the standard deviation, excludes the KL divergence term, and adopts the Clip-Higher~\cite{yu2025dapo} strategy.

\subsection{Metrics}

To evaluate the accuracy of different methods, we transform the generated responses into executable strategies for BATTLE-FIELD and execute them within our simulation environment. A strategy is considered correct if it can be successfully executed in the blue team environment, effectively defends against the red team’s attacks, and does not cause any disruption to the blue team’s system. The accuracy is then calculated as the ratio of correct responses to the total number.

\begin{figure}[!t]
\centering
\subfloat[]{\includegraphics[width=3.3in]{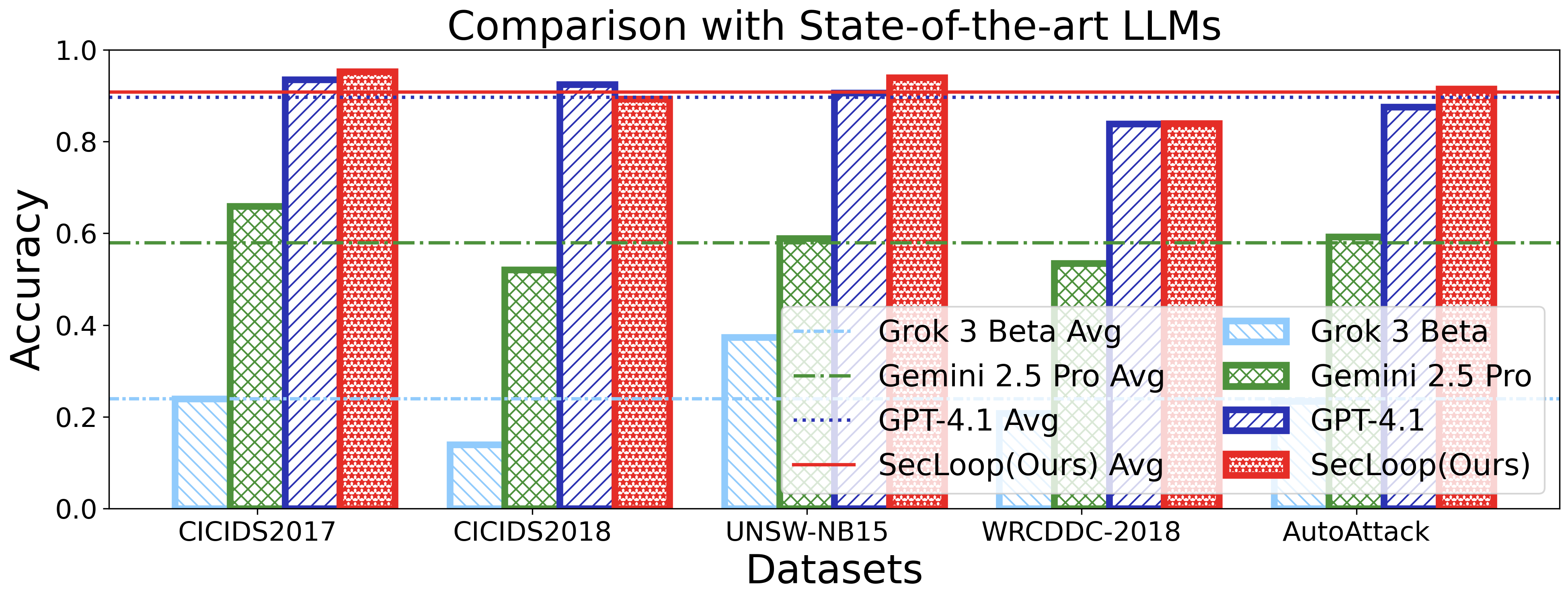}
\label{fig:5_first_case}}
\hfil
\subfloat[]{\includegraphics[width=3.3in]{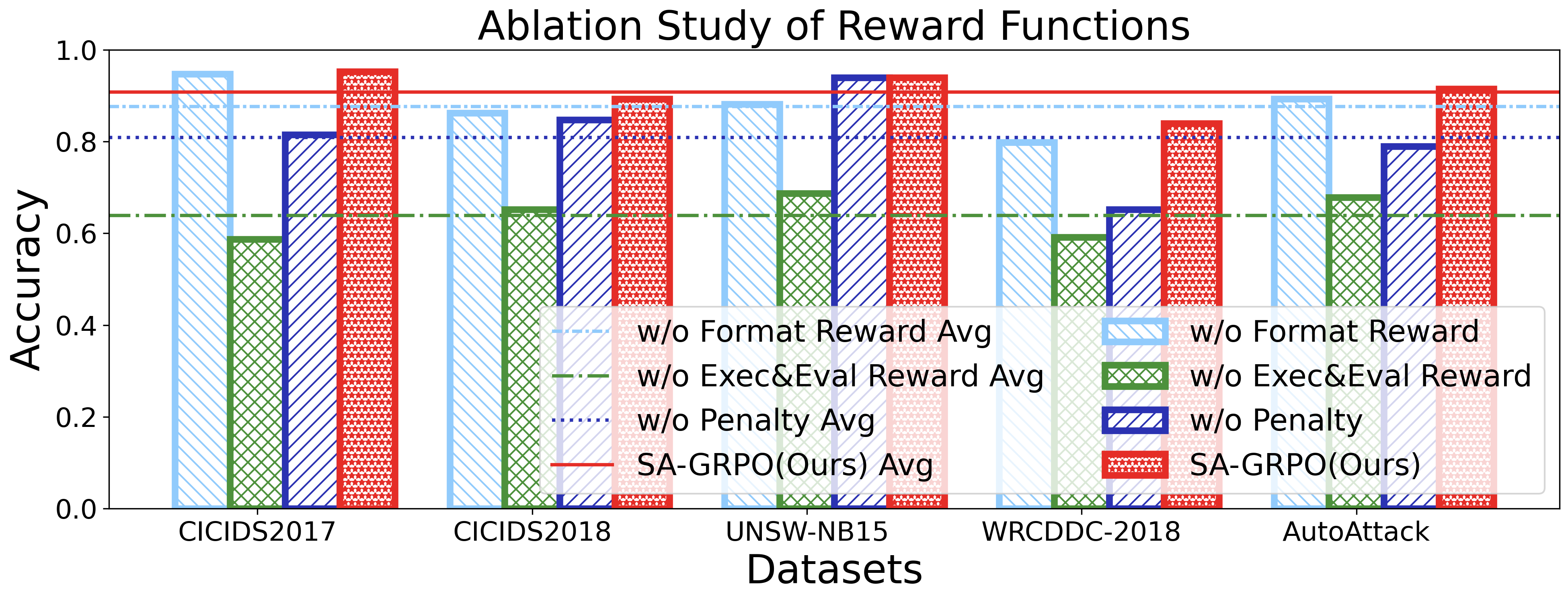}
\label{fig:5_second_case}}
\hfil
\subfloat[]{\includegraphics[width=3.3in]{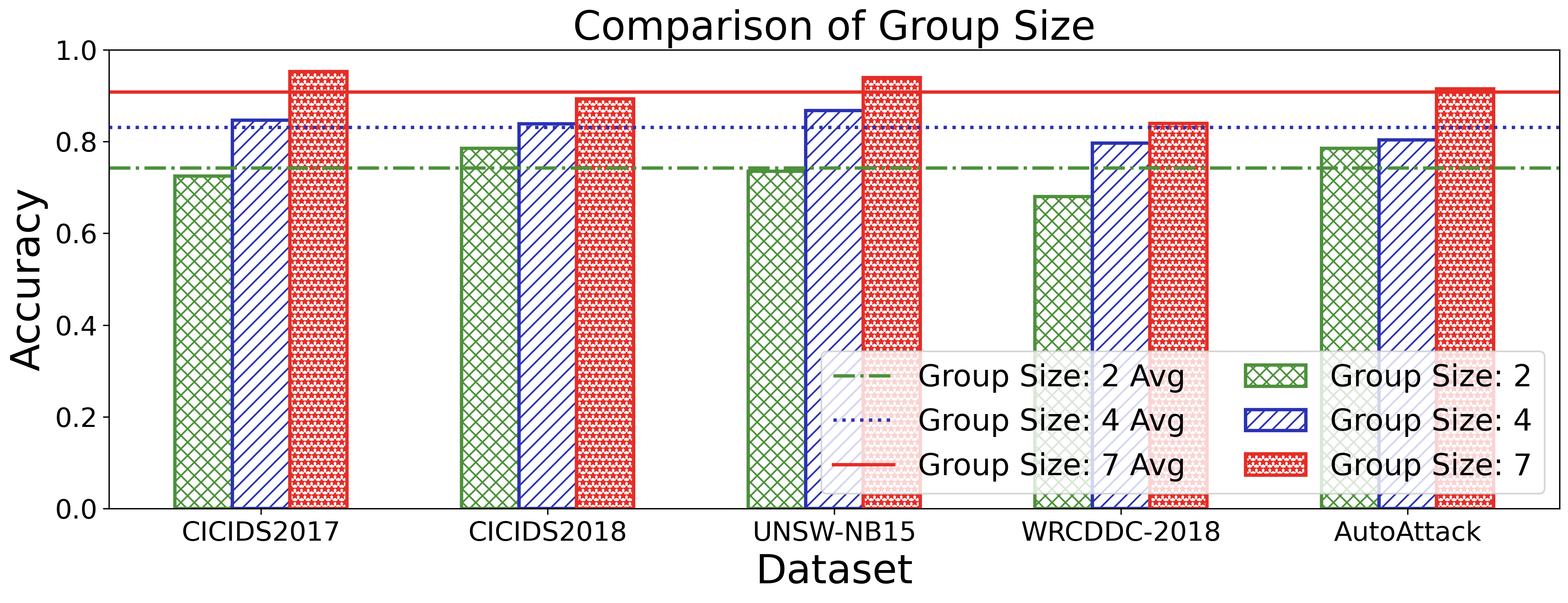}
\label{fig:5_third_case}}
\caption{Performance comparisons and ablation study of the proposed SecLoop and SA-GRPO. (a) compare the proposed SecLoop with state-of-the-art LLMs, (b) is the ablation study of reward functions of our SA-GRPO, and (c) demonstrates the impact of group size during training of our SA-GRPO.}
\label{fig:5}
\end{figure}

\subsection{Performance}
This section presents extensive experiments to evaluate the effectiveness of the proposed SecLoop and SA-GRPO methods, along with key observations and conclusions.

\ding{182} \textbf{Comparisons of Our Proposed SecLoop System:} 
As shown in Table \ref{tab:secloop}, we compare our proposed SecLoop in multiple dimensions, including security strategy generation, automated security orchestration, response, feedback, and ATT\&CK process completeness. It is evident that the SecLoop system achieves an end-to-end security automation loop. While existing systems demonstrate certain capabilities in strategy generation, many lack a fully integrated automated security orchestration process. Several systems do not support closed-loop feedback, limiting their capacity for iterative improvement and adaptation to new threats. In particular, we introduce a qualitative classification and statistical analysis method based on the MITRE ATT\&CK framework, which expands attack detection from a single dimension to tactical-level association analysis. Furthermore, our SecLoop system achieves the highest level of ATT\&CK process completeness.

We also compare the advancement of our system, SecLoop, in terms of functional coverage, decision-making intelligence, and adaptation capability.   In terms of functional coverage, most existing systems focus on specific stages of the security process, such as threat detection, policy generation, or policy optimization, whereas SecLoop achieves a complete closed loop from threat perception to policy self-optimization. Regarding decision-making intelligence, most existing security systems rely on manual configuration and rule-driven methods. In contrast, SecLoop automatically generates relevant scripts through the LLM to invoke security tools and employs SA-GRPO to achieve self-optimization of security decisions. Any manual intervention in these core modules would create a significant bottleneck, degrading the system into a traditional, slow-response security model. Compared with other systems, SecLoop can not only generate safety strategies but also invoke tools to execute strategies in different environments, which demonstrates a higher level of intelligence.   Finally, with respect to adaptation capability, compared to existing systems, SecLoop not only demonstrates stronger generalization across heterogeneous devices and dynamic environments but also adapts to various attack types, effectively reducing the cost of manual operations.

We compare the performance of our proposed SecLoop with two representative baselines on five benchmarks, as shown in Figure~\ref{fig:4}\subref{fig:4_first_case}. Experimental results show that SecLoop consistently outperforms existing methods across all test datasets, achieving average accuracy improvements of 41.6\% over xNIDS and 50.0\% over SAGE, respectively. Figure~\ref{fig:5}\subref{fig:5_first_case} presents a comparison of the performance of SecLoop with state-of-the-art LLMs, including Grok 3 Beta, Gemini 2.5 Pro Preview, and GPT-4.1, across five benchmark datasets. Notably, our SecLoop with 7B parameters consistently outperforms these large-parameter LLMs, achieving average accuracy improvements of 66.9\% over Grok 3 Beta, 32.9\% over Gemini 2.5 Pro Preview, and 1.2\% over GPT-4.1.

\ding{183} \textbf{Comparisons of Our Proposed SA-GRPO Algorithm:} 
To validate the effectiveness of our proposed SA-GRPO, we compare SA-GRPO with reinforcement learning algorithms, including DQN, DDQN, PPO, KTO, and GRPO, on five benchmarks. As illustrated in Figure~\ref{fig:4}\subref{fig:4_second_case}, SA-GRPO significantly outperforms other algorithms, showing average accuracy gains of 26.7\% over DQN, 22.7\% over DDQN, 18.6\% over PPO, 14.7\% over KTO, and 10.3\% over GRPO, respectively. Furthermore, we conduct an ablation study of SA-GRPO, as shown in Figure~\ref{fig:4}\subref{fig:4_third_case}. The results reveal substantial improvements in accuracy across all tasks after fine-tuning, with an average increase of 30.4\%. This demonstrates SA-GRPO's superior performance and stronger generalization capability against diverse attack scenarios.

Figure~\ref{fig:5}\subref{fig:5_second_case} presents an ablation study that investigates the impact of different rewards on the overall performance of SA-GRPO. We observe varying degrees of performance degradation when any of the rewards is removed. Specifically, the execution and evaluation rewards lead to the most significant impact on SA-GRPO, with an average decrease of 26.9\% compared to the original SA-GRPO. This indicates that the feedback from environments plays a critical role in guiding the model toward strategy generation. Disabling the penalty reward also results in a notable decline of 10.0\%, preventing destructive outputs and ensuring the security of our automated system. In contrast, the absence of the format reward causes the smallest performance drop at 3.1\%, because it mainly focuses on accelerating the training process of the model.

Figure~\ref{fig:5}\subref{fig:5_third_case} presents a sensitivity analysis on the impact of the group size, a key hyperparameter in our SA-GRPO algorithm. The results demonstrate a clear positive trend: as the group size increases, the model's average accuracy across all datasets steadily improves. This improvement is attributed to the core mechanism of group relative policy optimization; a larger group provides a wider and more diverse sample of candidate policies for each prompt, which allows for a more robust and lower-variance estimation of the advantage signal used for the policy update. However, this performance gain comes at the cost of increased computational overhead, as the number of generations and subsequent reward calculations scales with the group size. Considering this trade-off between performance and computational efficiency, we selected a group size of 7 for our primary experiments, as it delivered substantial accuracy gains while maintaining a manageable training cost.

\begin{table}[!t]
\centering
\caption{Performance Comparison of the Strategy on Different Hardware Platforms}
\label{tab:hardware_comparison}
\begin{tabular}{c|c|c|c|c}
\hline
\multicolumn{1}{c|}{\textbf{Platform}} & \textbf{\makecell{Tensor\\ Performance}} & \textbf{\makecell{VRAM\\ (MiB)}} & \textbf{\makecell{Time\\ cost (s)}} & \textbf{\makecell{Accuracy\\ (\%)}} \\
\hline
\textit{\begin{tabular}[c]{@{}c@{}}Simulation\\Environment\end{tabular}} & 1248 TOPS & 15410 & 0.74 & 92.71 \\
\hline
\textit{\begin{tabular}[c]{@{}c@{}}Edge\\Devices \end{tabular}} & 275 TOPS & 17868 & 48.19 & 91.35 \\
\hline
\end{tabular}
\end{table}

\ding{184} \textbf{Real-World Experiments:} 
To evaluate the deployment capability of SecLoop in resource-constrained environments, we conduct real-world tests on an NVIDIA Jetson edge device. The corresponding results are presented in Figures~\ref{fig:4}\subref{fig:4_fourth_case}\subref{fig:4_fifth_case}\subref{fig:4_sixth_case}. In Figure~\ref{fig:4}\subref{fig:4_fourth_case}, we deploy the model trained via SecLoop to the Jetson device and also implement the two representative baseline methods for comparative purposes. In the real-world test of SecLoop, we found that SecLoop still outperforms the other methods in terms of response accuracy under edge computing conditions, with average accuracy improvements of 41.6\% over xNIDS and 50.8\% over SAGE. This indicates that in the real-world environment, the performance of our SecLoop is similar to that of the simulation environment.

Figure~\ref{fig:4}\subref{fig:4_fifth_case} shows the performance of different reinforcement learning algorithms on the edge device, including PPO, KTO, GRPO, and our proposed SA-GRPO. The results reaffirm that SA-GRPO continues to outperform other algorithms even in resource-constrained environments, achieving average accuracy gains of 18.4\% over PPO, 16.4\% over KTO, and 11.2\% over GRPO. Moreover, the ablation study of SA-GRPO in the real-world situation demonstrates minor distinction from simulation environments, as shown in Figure~\ref{fig:4}\subref{fig:4_sixth_case}. These findings further prove SA-GRPO’s adaptability and high performance under resource-constrained conditions.

Table \ref{tab:hardware_comparison} presents a performance comparison of our proposed method on two different hardware platforms: a high-performance simulation environment and resource-constrained edge devices. The simulation environment, with a tensor performance of 1248 TOPS, utilizes 15410 MiB of VRAM, achieves the lowest time cost of 0.74 seconds, and an accuracy of 92.71\%. In contrast, the edge device, which operates under more limited computational resources with a tensor performance of 275 TOPS, has a measured peak memory usage of 17868 MiB, exhibits a higher time cost of 48.19 seconds, but still maintains a high accuracy of 91.35\%. Notably, the slightly higher memory consumption on edge devices is because of the unified memory system adopted by the edge platform and differences in optimization levels. This demonstrates that while the edge platform will increase time consumption, it maintains a high level of accuracy, highlighting the adaptability and robustness of our method across diverse hardware environments.

\begin{figure*}[!t]
\centering
\includegraphics[width=1.0\textwidth,trim=20 100 28 145,clip]{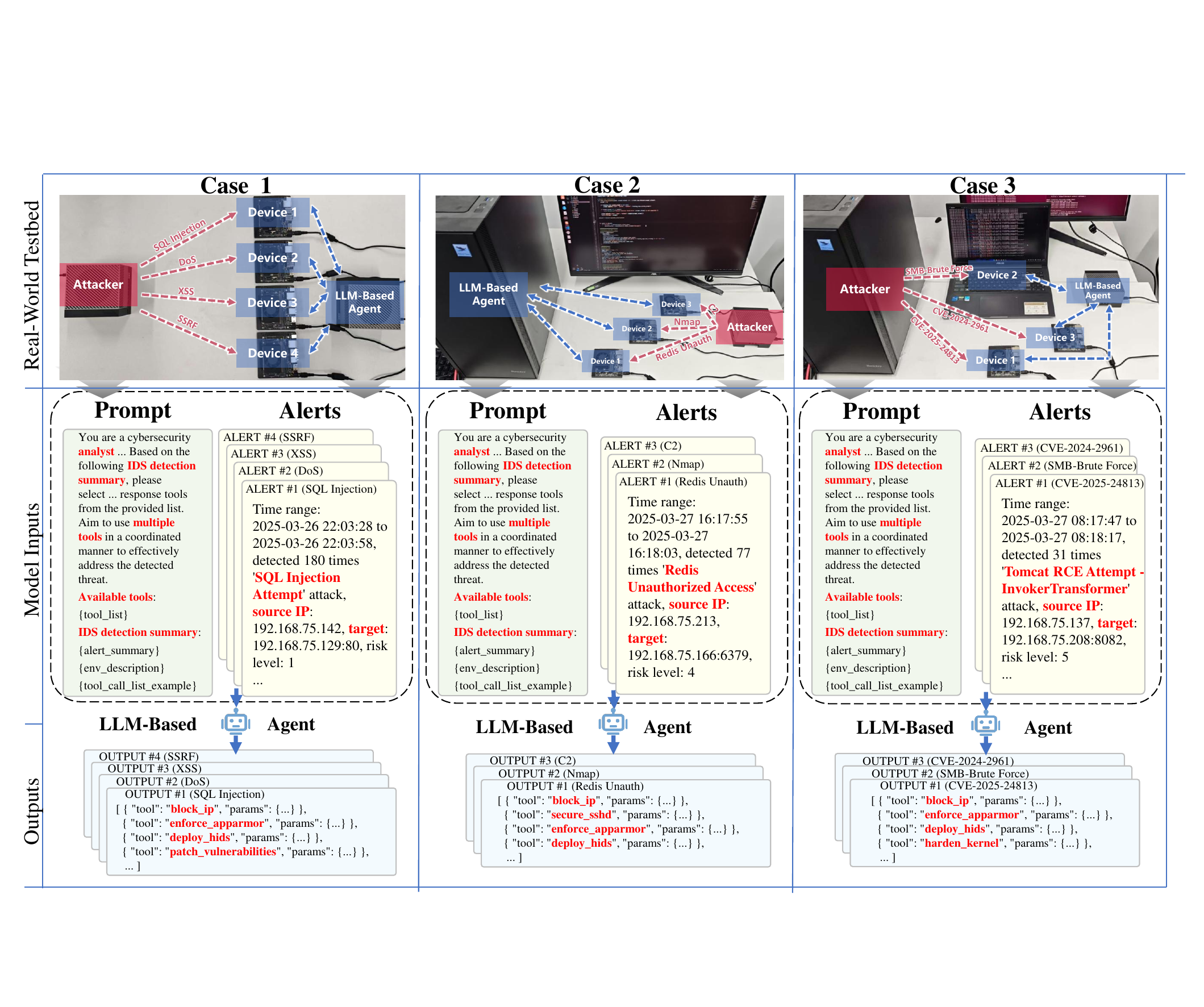}
\caption{Case study of the proposed SecLoop. At the top of this figure, three heterogeneous real-world environments composed of various embedded devices, a host, and a laptop. The workflow indicates that the attacker attempts to attack multiple devices. Subsequently, our LLM-based agent summarizes alerts and generates corresponding in-depth security strategies. Available security tools on the devices will be invoked to deal with the diverse attacks.}

\label{fig:6_case_study}
\end{figure*}

\subsection{Case Study}

As illustrated in Figure \ref{fig:6_case_study}, we visualize three case studies to demonstrate the effectiveness of our proposed SecLoop and SA-GRPO in three heterogeneous environments of real-world testbeds. For each scenario, we set up an attacker and multiple victim devices. The LLM-Based Agent collects the alerts sent by the victim devices and generates corresponding security strategies. These strategies guide the devices in activating security tools to mitigate diverse attacks.

In the first case, we used four edge devices (Jetson ORIN NX 16GB) as victim devices and a Jetson AGX Orin 64GB as the attacker. The LLM-based agent, deployed on the Jetson AGX Orin 64GB, generates strategies and orchestrates security. In this scenario, the attacker launches various attacks on the victim devices, including SQL injection, DoS, XSS, and SSRF. The collected alerts and prompts are fed into the model to generate executable security strategies. For example, on the victim devices, we deployed a web service with an SQL injection vulnerability that accepts unsanitized user input. Sensitive information is sent back to the attacker’s platform through cross-protocol communication. When the model receives the ``SQL Injection Attempt'' alert, it outputs strategies like blocking IP and restraining other ongoing processes, effectively defending the system against the attack.

In the second case, we deployed three Jetson ORIN NX edge devices as victims and a Jetson AGX Orin edge device as the attacker. The LLM-based agent, running on a host machine, handles strategy generation and security orchestration. The attacker targets the three victim devices with attacks such as Command\&Control, Nmap Scanning, and Redis Unauthorized Write. Attack validation occurs in three stages: 1) establishing a reverse TCP connection with netcat to test file writeability, 2) executing ``whoami'' to verify privilege escalation, 3) using ``curl'' to trigger the WebShell and verify attack chain integrity. When the model detects the ``Redis Unauthorized Access'' alert, it generates defense strategies such as blocking IP, enhancing SSH service security, and restraining other ongoing processes to mitigate the attack.

In the third case, we deployed two Jetson ORIN NX devices and a laptop as victim devices, with a host acting as the attacker. The LLM-based agent, deployed on a Jetson AGX Orin, generates strategies and orchestrates security. The attacker targets the victims with different attacks, including CVE-2025-24813, SMB-Brute Force, and CVE-2024-2961. The alerts and prompts are processed by the model to generate executable security strategies. Attack verification occurs in three stages: 1) monitoring Tomcat logs to confirm file path; 2) using the JMX protocol to validate malicious class loading; 3) using the ps command to detect injected commands. Upon receiving the ``Tomcat RCE Attempt - InvokerTransformer'' alert, the model outputs strategies like blocking IP, restraining other ongoing processes, and hardening kernel configuration to mitigate the attack.

\section{Discussions}
\label{5dis}
So far, we have shown the superiority of our proposed SecLoop and SA-GRPO over five benchmarks. In this section, we take the next step to highlight some interesting observations and future work, including the adaptivity, generalization, and self-evolution of the proposed SecLoop and SA-GRPO.
\subsection{Adaptivity of Our Proposed SA-GRPO to Various Attacks}
For the ATT\&CK process, we have implemented the automated generation and environment replay of over 20 different attacks as listed in Table \ref{tab:MITRE ATTCK}. The mitigation tools and execution methods required for these attacks vary significantly. Our proposed SA-GRPO algorithm is capable of identifying the appropriate tools for various attacks and automatically invoking them, demonstrating its practicality in real-world environments. This approach advances automated network security mitigation and reduces the cost of manual operations.

\subsection{Generalization of Our Proposed SA-GRPO to Heterogeneous Environments}
Communication security scenarios often involve heterogeneous devices and dynamic environments, which offer high demands on the model's generalization capability. We test the SA-GRPO model in three different heterogeneous real-world environments and found that it achieved an accuracy rate of over 91\% as shown in Figure~\ref{fig:4}\subref{fig:4_fifth_case}, demonstrating the model's potential for practical deployment.

\subsection{Self-Evolving of Our Proposed SecLoop}
Due to the dynamic nature of environments and software version updates, tool invocation presents significant challenges, requiring continuous updates and evolution of the model. In the future, we plan to conduct in-depth research into the self-evolution mechanism of LLMs, leveraging online feedback from actual environments to drive the self-adaptation of the SecLoop system’s security capabilities and minimize potential adverse impacts such as over-defense. Additionally, we will explore the collaboration between large and small models to reduce system resource consumption and latency. By implementing dynamic task allocation and knowledge distillation techniques, we aim to significantly reduce resource consumption and latency, thereby enhancing the system's suitability for resource-constrained deployment scenarios without compromising security effectiveness.

\subsection{Fault Tolerance and Proactive Threat Discovery}
To enhance the system for production-grade deployment, our future work will also focus on developing robust fault tolerance and proactive defense capabilities. We plan to explore a ``Guardian Model'' framework, where an independent monitoring model performs real-time risk assessment on the strategies generated by the main agent, intervening to prevent high-risk actions. Concurrently, we will investigate integrating formal verification methods to logically validate strategies against predefined security invariants before execution. Furthermore, we aim to evolve SecLoop from a reactive system to one capable of proactive threat discovery by incorporating advanced techniques, such as program analysis and fuzz testing, to autonomously identify potential vulnerabilities.

\section{Conclusion}
\label{6con}
In this work, we address the pressing security challenges posed by the dynamic, open, and heterogeneous nature of 6G Zero-Touch Networks by introducing SecLoop and SA-GRPO. SecLoop provides a fully automated, end-to-end security automation framework that integrates LLMs across the entire lifecycle of strategy generation, orchestration, execution, and feedback, enabling intelligent and adaptive defense in real-world adversarial environments. Building on this foundation, SA-GRPO serves as a novel group-based reinforcement learning algorithm that optimizes security strategies without relying on high-quality labeled data, leveraging parallel execution and feedback-driven refinement. Experimental results demonstrate the effectiveness of our approach in heterogeneous network scenarios. We believe that SecLoop and SA-GRPO offer a practical and extensible foundation for intelligent security automation in ZTNs. In the future, we plan to explore online self-evolution of the proposed SecLoop system.

\appendices
\section{Hyperparameter Configurations}
\label{Appendix Hyperparameter}

To ensure the reproducibility of our work, this section provides a detailed breakdown of the hyperparameter configurations used for training our model with the SA-GRPO algorithm. Our experiments are based on the \texttt{Qwen2.5-7B-Instruct} model. We utilized the DeepSpeed ZeRO-2 strategy for distributed training across 8 GPUs, optimizing for computational efficiency with a \texttt{bfloat16} mixed-precision setup.

Table \ref{tab:hyperparams_comparison} summarizes these key hyperparameters and algorithm-specific configurations used in our model training and optimization strategy.

\begin{table}[!t]
\centering
\caption{Key Hyperparameters, Descriptions, and Configurations}
\label{tab:hyperparams_comparison}
\begin{tabular}{m{1.9cm}<{\centering}|m{4cm}|c}
\hline
\textbf{\makecell{Hyper\\Parameter}} & \textbf{Description} & \textbf{Value} \\
\hline
$G$ & Number of response samples generated per query for group-relative advantage estimation; controls diversity and stability of policy updates. & 7 \\
\hline
$\varepsilon_\mathrm{low}$ & Lower bound for clipping the probability ratio $\gamma_{i,t}(\theta)$ in the surrogate objective; avoids overly conservative updates. & 0.2 \\
\hline
$\varepsilon_\mathrm{high}$ & Upper bound for clipping $\gamma_{i,t}(\theta)$; limits aggressive updates to promote stability and prevent large policy shifts. & 0.28 \\
\hline
reward\_weights & Weights for each reward function. & 1 \\
\hline
$\alpha$ & Controls the mix between the current policy and the previous reference policy during updates. & 0.6 \\
\hline
$\beta$ & Coefficient for KL divergence penalty, controlling the regularization against the reference policy for stable fine-tuning. & 0 \\
\hline
Learning Rate & Step size for updating policy parameters during optimization. & $3\mathrm{e}{-6}$ \\
\hline
ref\_model\_sync\_ steps & Determines how frequently the current policy is synchronized with the reference policy. & 512 \\
\hline
torch\_dtype & Mixed-precision training using bfloat16 for computational efficiency. & bfloat16 \\
\hline
num\_train\_epochs & Total number of training epochs. & 1 \\
\hline
per\_device\_train\_ batch\_size & Batch size per GPU. & 4 \\
\hline
gradient\_accu-mulation\_steps & Gradient accumulation steps. & 1 \\
\hline
\end{tabular}
\end{table}

\section{Time Complexity Analysis}
\label{Appendix Time Complexity}

In this section, we provide a formal theoretical analysis of the SecLoop framework's complexity in terms of time, messages, and communication rounds. The analysis is presented for the two primary operational phases: the training phase and the inference (deployment) phase. We define the following variables for our analysis: $B$ denotes the batch size of prompts, $G$ is the group size (number of generations per prompt), $L_p$ is the length of the input prompt sequence, $L_c$ is the length of the generated completion (policy) sequence, $d_{\text{model}}$ is the hidden dimension of the model, $P$ is the total number of model parameters, $N_{\text{env}}$ is the number of parallel BATTLE-FIELD environments, and $T_{\text{sim}}$ is the time required for a single BATTLE-FIELD simulation.
\subsection{Complexity Analysis of the Training Phase}
During the training phase, the system optimizes the model's policy through iterative steps. Each training step comprises three core stages: policy generation, parallel evaluation, and policy update.

\textbf{1) Time Complexity of Policy Generation:}
The objective of this stage is to generate $G$ candidate policies for each of the $B$ prompts. The time complexity is derived from the first principles of the Transformer architecture.

First, the complexity of a single forward pass for a sequence of length $L$ is dominated by the self-attention mechanism ($O(L^2 \cdot d_{\text{model}})$) and the feed-forward network ($O(L \cdot d_{\text{model}}^2)$). Thus, the total complexity is given by:
\begin{equation}
    T_{\text{single\_pass}}(L) = O(L^2 \cdot d_{\text{model}} + L \cdot d_{\text{model}}^2).
\end{equation}

During autoregressive generation with a KV cache, a policy of length $L_c$ is generated from a prompt of length $L_p$. This involves a \textit{Prefill} operation on the prompt to compute and cache KV vectors, with a complexity of $T_{\text{prefill}} = T_{\text{single\_pass}}(L_p)$. Subsequently, a \textit{Decoding} loop generates tokens sequentially. The complexity to generate the $i$-th token is approximately $O((L_p + i) \cdot d_{\text{model}}^2)$. Therefore, the total decoding complexity for $L_c$ tokens is the sum over all generated tokens:
\begin{equation}
    \sum_{i=1}^{L_c} O((L_p + i - 1) \cdot d_{\text{model}}^2),
\end{equation}
which is approximately:
\begin{equation}
    O(L_c \cdot (L_p + L_c) \cdot d_{\text{model}}^2).
\end{equation}

For a total batch size of $N = B \cdot G$, we extend the above complexity to the entire batch. This yields the total time complexity formula for the policy generation stage:
\begin{equation}
\begin{aligned}
    T_{\text{generation}} = & O\Big( B \cdot G \cdot \big(L_p^2 \cdot d_{\text{model}} + L_p \cdot d_{\text{model}}^2 \\
    & + L_c \cdot (L_p + L_c) \cdot d_{\text{model}}^2\big)\Big).
\end{aligned}
\end{equation}

\textbf{2) Complexity of Parallel Evaluation in BATTLE-FIELD:}
The $B \times G$ generated policies are distributed among $N_{\text{env}}$ parallel BATTLE-FIELD environments for evaluation.
\begin{itemize}
    \item \textbf{Time Complexity:} For a single simulation, the time complexity is:
\begin{equation}
O(T_{\text{sim}}\times\left\lceil\frac{B\times G}{N_{\text{env}}}\right\rceil).
\end{equation}
    \item \textbf{Messages and Rounds:} This stage constitutes the main external communication overhead. Within one training step, there is one core interaction: the trainer sends all policies to the BATTLE-FIELD server, and the server returns the corresponding scores. The number of logical messages is proportional to the total number of policies, i.e., $O(B \cdot G)$.
\end{itemize}

\textbf{3) Time Complexity of Policy Update:}
This stage is a standard backpropagation process, with a computational cost proportional to the forward pass. The complexity of one backpropagation pass is proportional to the product of the batch size, sequence length, and the number of model parameters. For a total batch size of $N = B \cdot G$ and a total sequence length of $L = L_p + L_c$, the time complexity for the policy update stage is $T_{\text{update}} = O(B \cdot G \cdot (L_p + L_c) \cdot P)$.
\subsection{Complexity Analysis of the Inference (Deployment) Phase}
During deployment for autonomous response, the analysis focuses on the end-to-end response time for a single event.

\textbf{1) Time Complexity:}
The total response time $T_{\text{response}}$ is the sum of three sequential stages: $T_{\text{response}} = T_{\text{infer}} + T_{\text{comm}} + T_{\text{exec}}$.
\begin{itemize}
    \item \textbf{Policy Generation ($T_{\text{infer}}$):} For a single inference pass where $B=1$ and $G=1$, the generation time is:
\begin{equation}
\begin{aligned}
T_{\text{infer}} = & O(L_{\text{alert}}^2 \cdot d_{\text{model}} + L_{\text{alert}} \cdot d_{\text{model}}^2 + \\
    & L_{\text{policy}} \cdot (L_{\text{alert}} + L_{\text{policy}}) \cdot d_{\text{model}}^2).
\end{aligned}
\end{equation}
    \item \textbf{Communication Latency ($T_{\text{comm}}$):} The delay from sending the policy over the network.
    \item \textbf{Execution Latency ($T_{\text{exec}}$):} The time required for security tools to execute on the target system.
\end{itemize}

\textbf{2) Messages and Rounds:}
The communication model during inference is a fixed 2-round interaction protocol for a single, decisive action:
\begin{itemize}
    \item \textbf{Round 1:} The sensor/IDS sends 1 alert message to the LLM Agent.
    \item \textbf{Round 2:} The LLM Agent sends 1 policy directive to the execution endpoint.
\end{itemize}
This framework is extensible to handle complex incidents requiring iterative reasoning. In such cases, the protocol can be expanded to $2n$ rounds, where $n$ is the number of steps in the response plan. After each action, the execution result serves as a new observation, initiating a subsequent 2-round cycle for the next-step decision.

\bibliographystyle{IEEEtran}
\bibliography{reference.bib}

\begin{IEEEbiography}[{\includegraphics[width=1in,height=1.25in,clip,keepaspectratio]{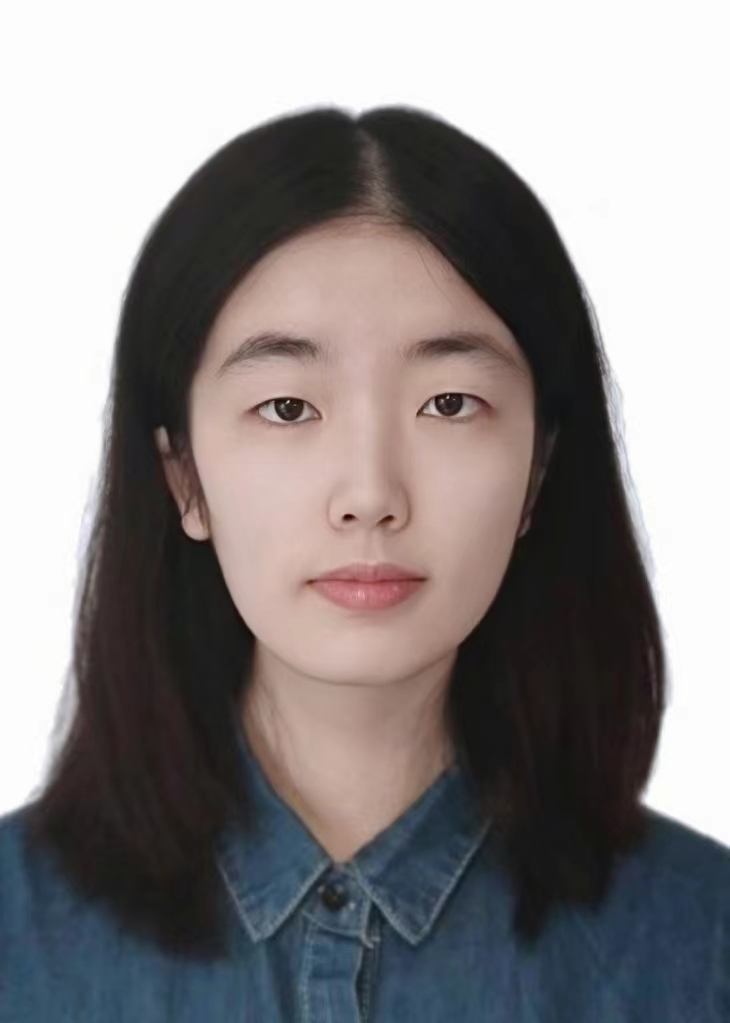}}]{Xinye Cao}
(Graduate Student Member, IEEE) received the B.E. degree in electronic and information engineering from Central China Normal University (CCNU), Wuhan, China, in 2020. She is currently pursuing the Ph.D. degree in cyberspace security with the National Engineering Research Center for Mobile Network Technologies, Beijing University of Posts and Telecommunications (BUPT). Her research interests include large AI model for future wireless communication systems and wireless communication security.
\end{IEEEbiography}

\begin{IEEEbiography}[{\includegraphics[width=1in,height=1.25in,clip,keepaspectratio]{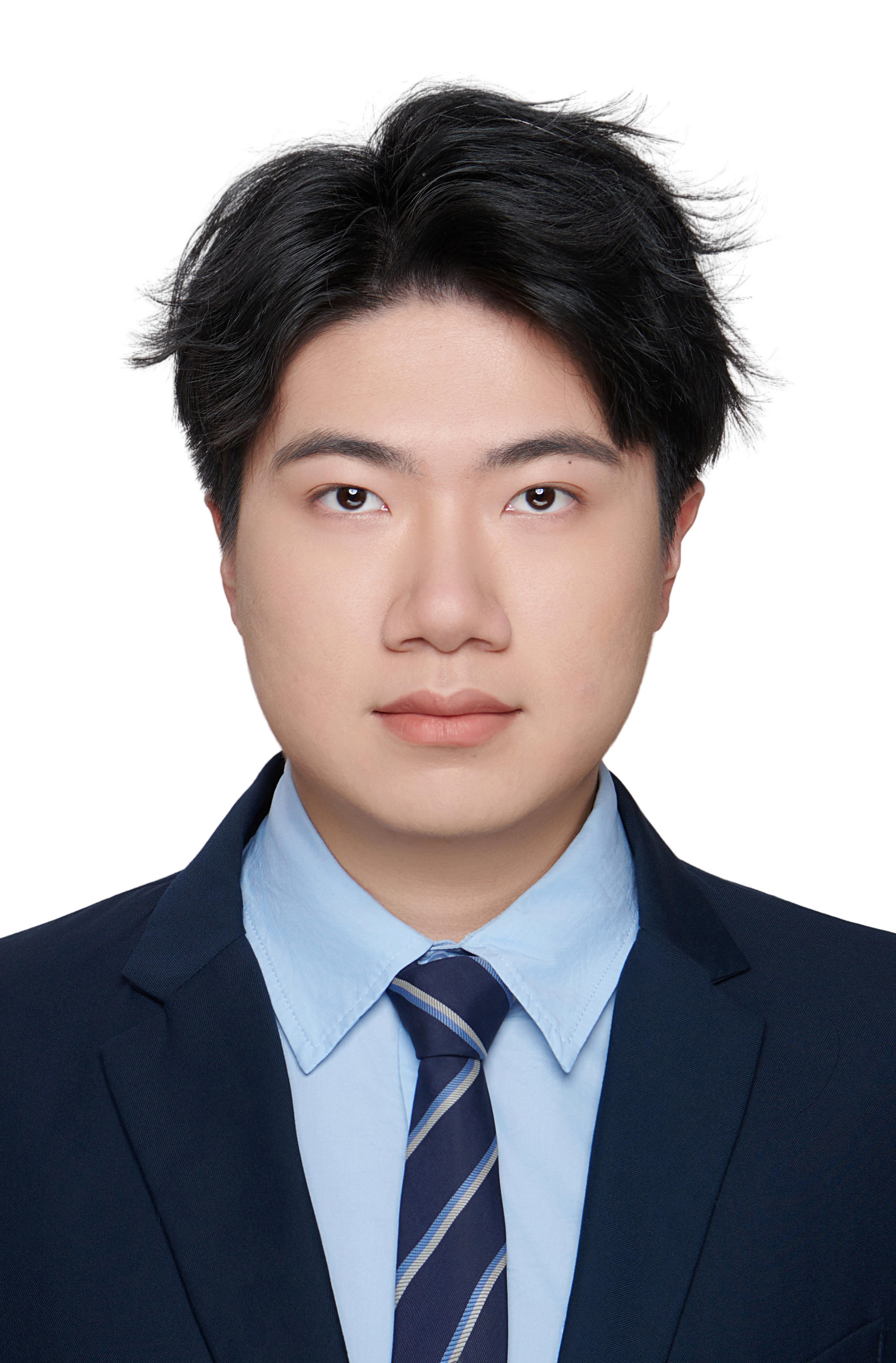}}]{Yihan Lin}
is currently pursuing the bachelor’s degree with the Beijing University of Posts and Telecommunications (BUPT), Beijing, China. He is working as a Research Intern with the National Engineering Research Center for Mobile Network Technologies, BUPT. His research interests include wireless communications security and software security.
\end{IEEEbiography}

\begin{IEEEbiography}[{\includegraphics[width=1in,height=1.25in,clip,keepaspectratio]{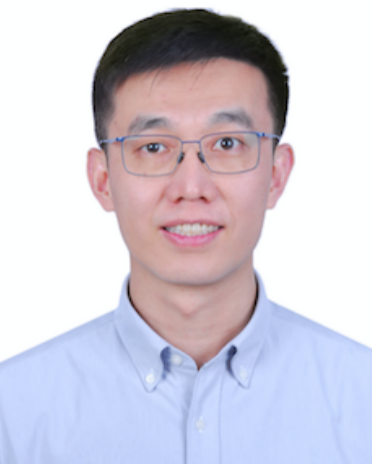}}]{Guoshun Nan}
(Member, IEEE) is a full professor at National Engineering Research Center for Mobile Network Technologies, Beijing University of Posts and Telecommunications, Beijing, China. He has broad interest in multimodal learning, large language models, and 6G network security, and has published more than 30 papers in top-tier journals and conferences including IEEE TPAMI, IEEE JSAC, IEEE TMC, IEEE TIFS, NeurIPS, ICML, CVPR, ICCV, ACL, etc. He also serves as a reviewer for these communities.
\end{IEEEbiography}

\begin{IEEEbiography}[{\includegraphics[width=1in,height=1.25in,clip,keepaspectratio]{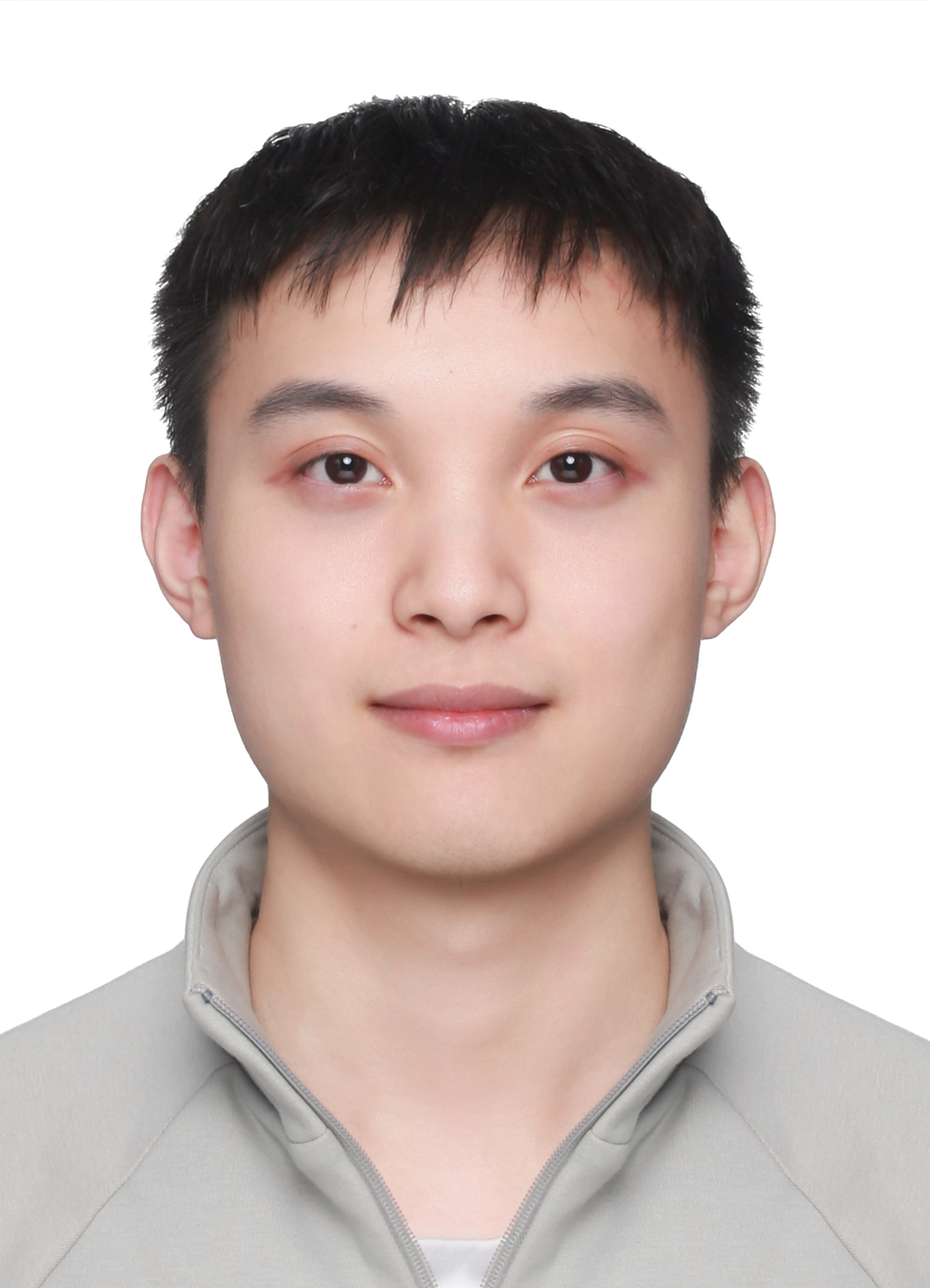}}]{Qinchuan Zhou}
is currently pursuing the bachelor’s degree with the Beijing University of Posts and Telecommunications (BUPT), Beijing, China. He is working as a Research Intern with the National Engineering Research Center for Mobile Network Technologies, BUPT. His research interests encompass computer vision and large language model based multi-agent architectures.
\end{IEEEbiography}

\begin{IEEEbiography}[{\includegraphics[width=1in,height=1.25in,clip,keepaspectratio]{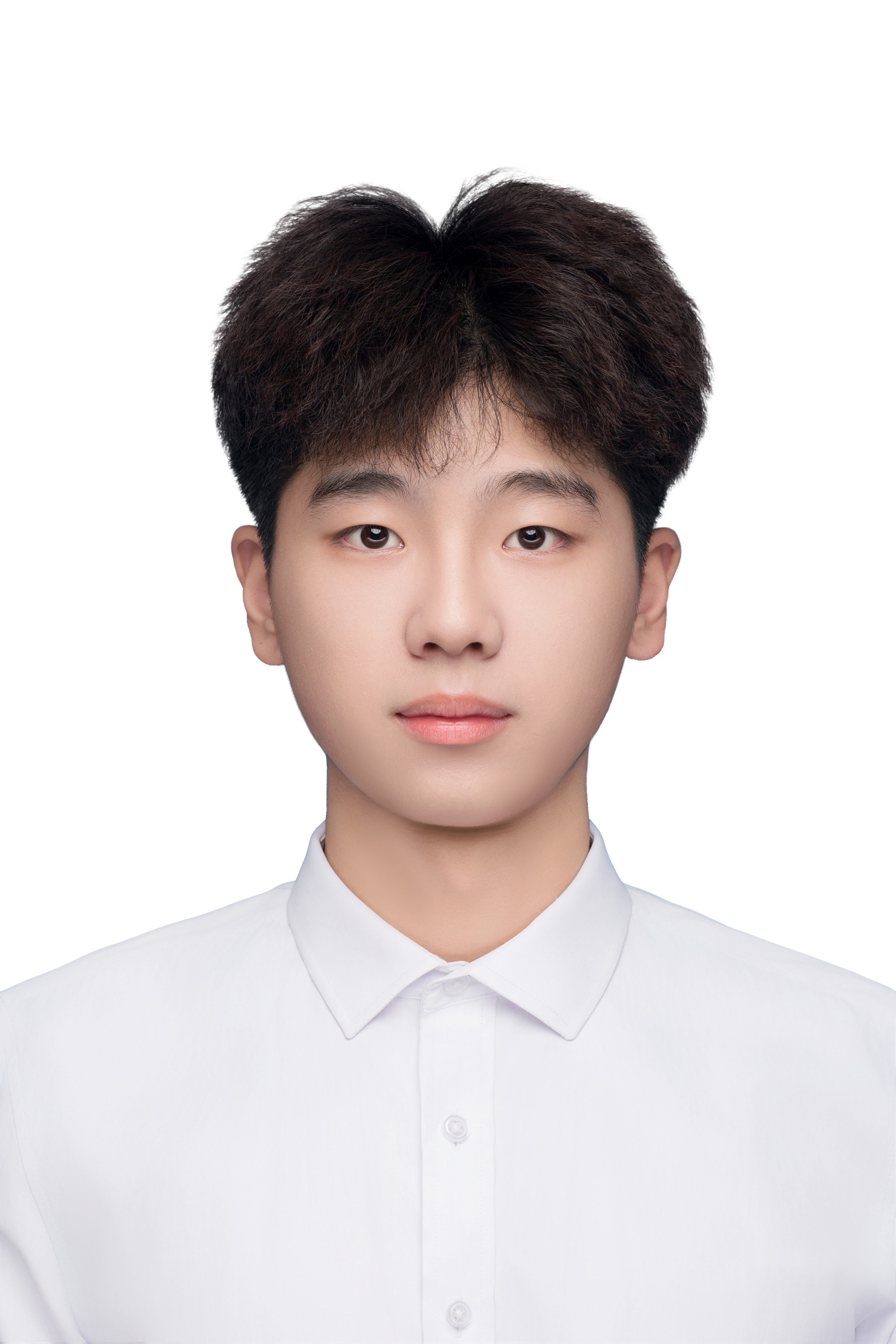}}]{Yuhang Luo}
is currently pursuing the bachelor’s degree with Beijing University of Posts and Telecommunications (BUPT), Beijing, China. He is working as a Research Intern with the National Engineering Research Center for Mobile Network Technologies, BUPT. His research interests include penetration testing, vulnerability mining, WEB security, and firewall security.
\end{IEEEbiography}

\begin{IEEEbiography}[{\includegraphics[width=1in,height=1.25in,clip,keepaspectratio]{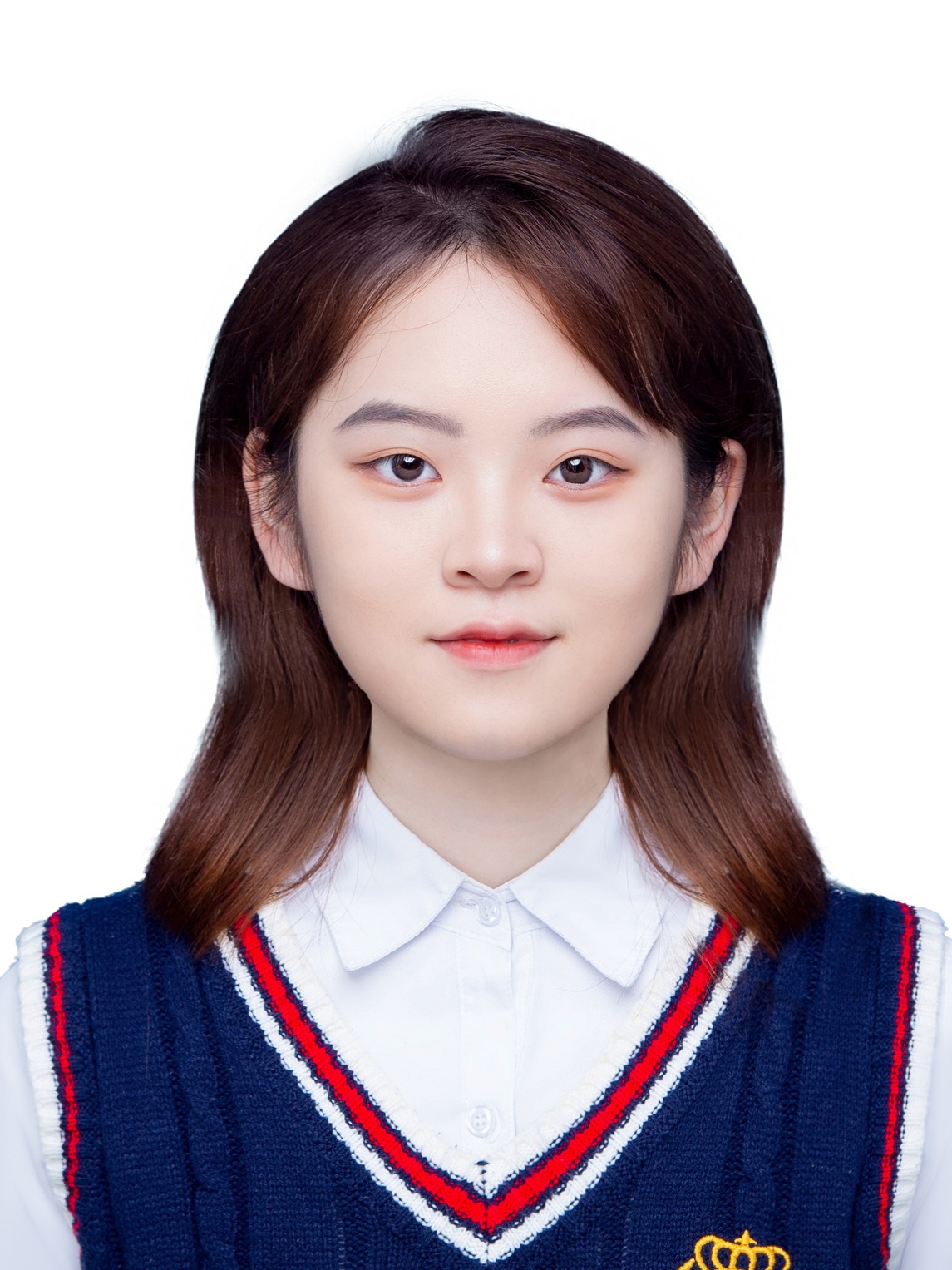}}]{Yurui Gao}
is currently pursuing the bachelor's degree with Beijing University of Posts and Telecommunications (BUPT), Beijing, China. She is working as a Research Intern with the National Engineering Research Center for Mobile Network Technologies, BUPT. Her research interests include large language model security, software security, and program analysis.
\end{IEEEbiography}

\begin{IEEEbiography}[{\includegraphics[width=1in,height=1.25in,clip,keepaspectratio]{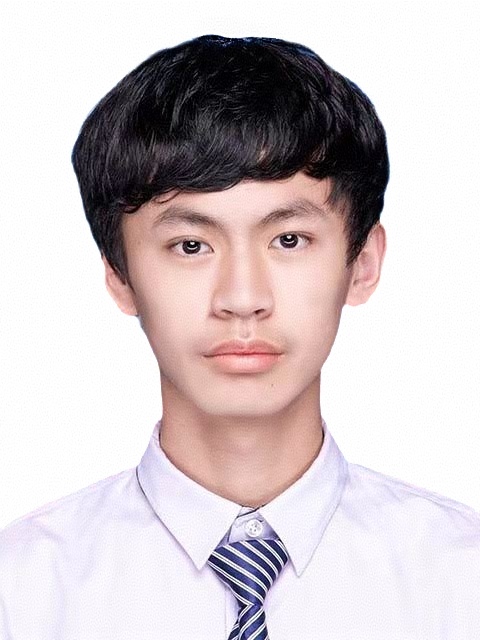}}]{Zeliang Zhang}
is currently pursuing the bachelor’s degree with Beijing University of Posts and Telecommunications (BUPT), Beijing, China. He is working as a Research Intern with the National Engineering Research Center for Mobile Network Technologies, BUPT. His research interests include cybersecurity and large language model.
\end{IEEEbiography}

\begin{IEEEbiography}[{\includegraphics[width=1in,height=1.25in,clip,keepaspectratio]{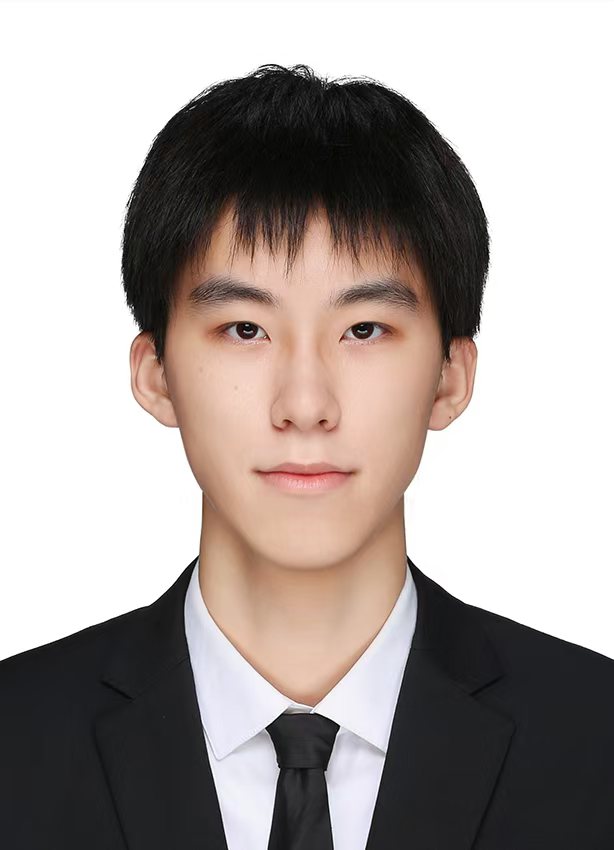}}]
{Haolang Lu}
(Graduate Student Member, IEEE) received the B.E. degree in cyberspace security from Beijing University of Posts and Telecommunications (BUPT), Beijing, China, in 2024. He is currently pursuing the Ph.D. degree with the Graduate College for Engineers, BUPT. His research interests include artificial intelligence security and interpretability.
\end{IEEEbiography}

\begin{IEEEbiography}[{\includegraphics[width=1in,height=1.25in,clip,keepaspectratio]{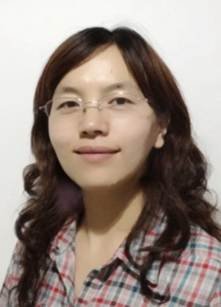}}]{Qimei Cui}
(M’09–SM’15) received the B.E. and M.S. degrees in electronic engineering from Hunan University, Changsha, China, in 2000 and 2003, respectively, and the Ph.D. degree in information and communications engineering from the Beijing University of Posts and Telecommunications (BUPT), Beijing, China, in 2006. She has been a Full Professor with the School of Information and Communication Engineering, BUPT, since 2014. She was a visiting Professor with the Department of Electronic Engineering, University of Notre Dame, IN, USA, in 2016. Her research interests include B5G/6G wireless communications, mobile computing and IoT. She serves as a Technical Program Chair of the APCC 2018, and a Track Chair of IEEE/CIC ICCC 2018, and a Workshop Chair of WPMC 2016. She also serves as a Technical Program Committee Member of several international conferences, such as the IEEE ICC, the IEEE WCNC, the IEEE PIMRC, the IEEE ICCC, the WCSP 2013, and the IEEE ISCIT 2012. She won the Best Paper Award at the IEEE ISCIT 2012, the IEEE WCNC 2014, and the WCSP 2019, and the Honorable Mention Demo Award at the ACM MobiCom 2009, and the Young Scientist Award at the URSI GASS 2014. She serves as Editor of SCIENCE CHINA Information Science , and Guest Editor of the EURASIP Journal on Wireless Communications and Networking and International Journal of Distributed Sensor Networks and Journal of Computer Networks and Comm.
\end{IEEEbiography}

\begin{IEEEbiography}[{\includegraphics[width=1in,height=1.25in,clip,keepaspectratio]{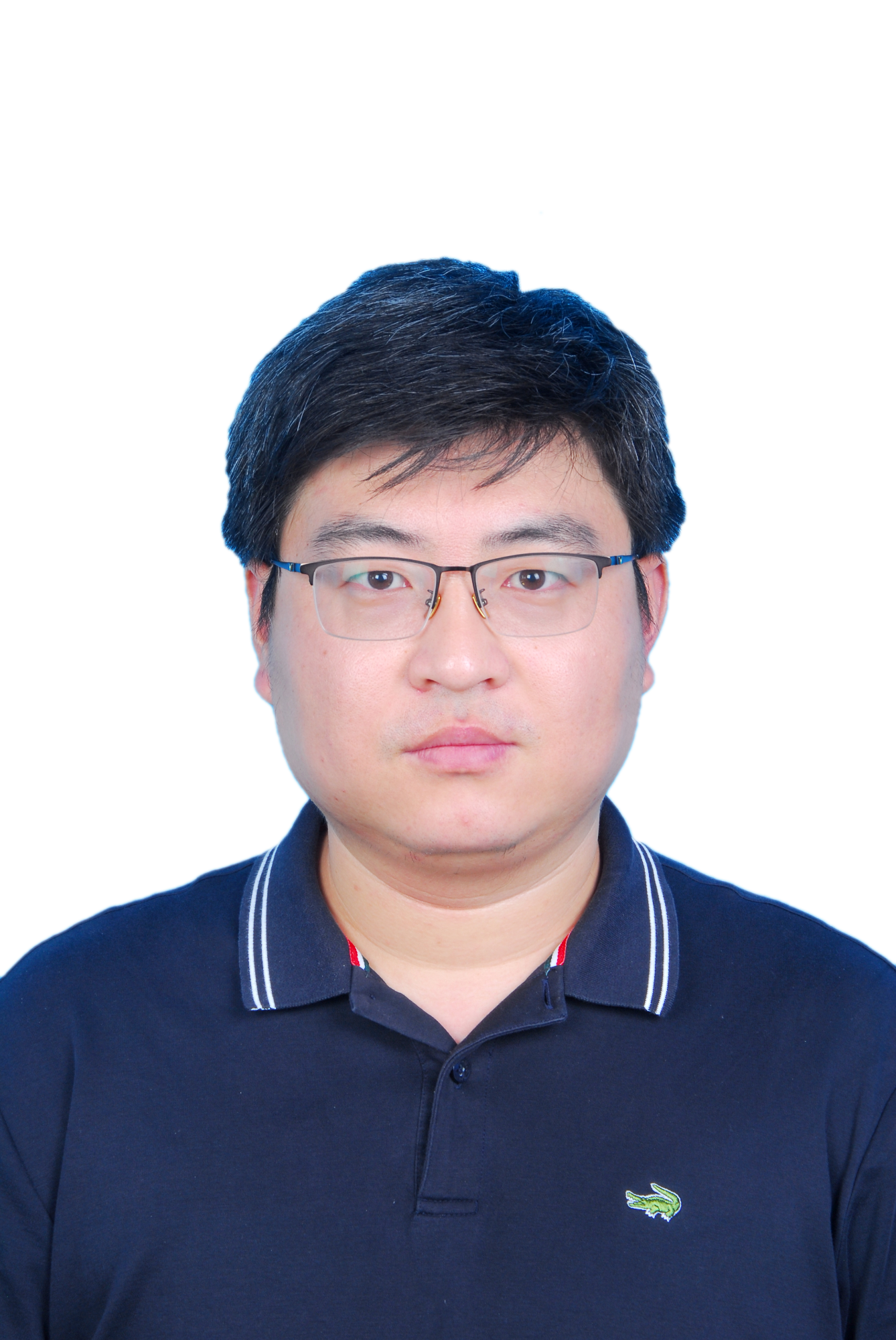}}]{Yanzhao Hou}
(Member, IEEE) received the Ph.D. degree from the Beijing University of Posts and Telecommunications (BUPT), Beijing, China, in 2014. He is currently with the National Engineering Research Center for Mobile Network Technologies, BUPT. His current research interests include wireless federated learning, AI-driven wireless communications and trial systems. He received the Best Demo Award in IEEE APCC2018.
\end{IEEEbiography}

\begin{IEEEbiography}[{\includegraphics[width=1in,height=1.25in,clip,keepaspectratio]{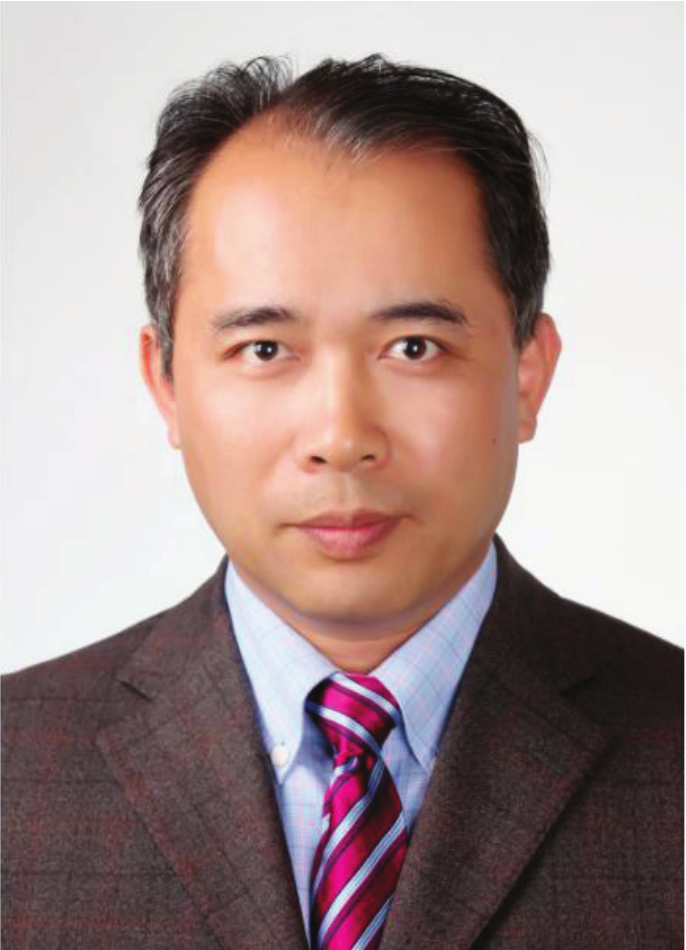}}]{Xiaofeng Tao}
received the B.S. degree in electrical engineering from Xi’an Jiaotong University, Xi’an, China, in 1993, and the M.S. and Ph.D. degrees in telecommunication engineering from Beijing University of Posts and Telecommunications(BUPT), Beijing, China, in 1999 and 2002, respectively. He is a Professor in BUPT, a Fellow of the Institution of Engineering and Technology, and Chair of the IEEE ComSoc Beijing Chapter. He has authored or co-authored over 200 papers and three books in wireless communication areas. He focuses on 5G/B5G research.
\end{IEEEbiography}

\begin{IEEEbiography}[{\includegraphics[width=1in,height=1.25in,keepaspectratio]{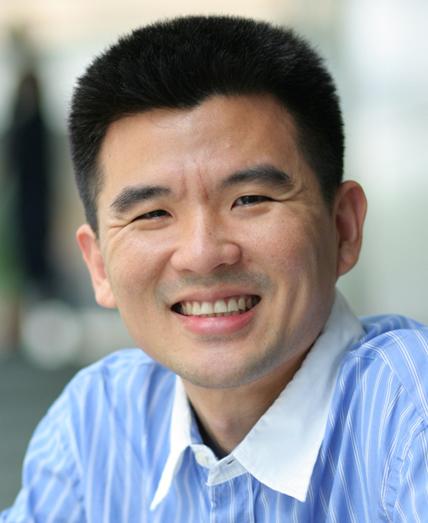}}]
{Tony Q.S. Quek}(S'98-M'08-SM'12-F'18) received the B.E.\ and M.E.\ degrees in electrical and electronics engineering from the Tokyo Institute of Technology in 1998 and 2000, respectively, and the Ph.D.\ degree in electrical engineering and computer science from the Massachusetts Institute of Technology in 2008. Currently, he is the Associate Provost (AI \& Digital Innovation) and Cheng Tsang Man Chair Professor with Singapore University of Technology and Design (SUTD). He also serves as the Director of the Future Communications R\&D Programme, and the ST Engineering Distinguished Professor. He is a co-founder of Silence Laboratories and NeuroRAN. His current research topics include wireless communications and networking, network intelligence, non-terrestrial networks, open radio access network, AI-RAN, and 6G.

Dr.\ Quek was honored with the 2008 Philip Yeo Prize for Outstanding Achievement in Research, the 2012 IEEE William R. Bennett Prize, the 2015 SUTD Outstanding Education Awards -- Excellence in Research, the 2016 IEEE Signal Processing Society Young Author Best Paper Award, the 2017 CTTC Early Achievement Award, the 2017 IEEE ComSoc AP Outstanding Paper Award, the 2020 IEEE Communications Society Young Author Best Paper Award, the 2020 IEEE Stephen O. Rice Prize, the 2020 Nokia Visiting Professor, the 2022 IEEE Signal Processing Society Best Paper Award, the 2024 IIT Bombay International Award For Excellence in Research in Engineering and Technology, the IEEE Communications Society WTC Recognition Award 2024, and the Public Administration Medal (Bronze). He is an IEEE Fellow, a WWRF Fellow, an AIIA Fellow, and a Fellow of the Academy of Engineering Singapore.
\end{IEEEbiography}

\vfill

\end{document}